\documentclass{aip-cp}

\usepackage[numbers]{natbib}
\usepackage{rotating}
\usepackage{graphicx}
\usepackage{amssymb}
\usepackage{xcolor}

\begin{document}

\title{Applicability of Relativistic Point-Coupling Models to Neutron Star Physics}

\author[aff1]{Bao Yuan Sun\corref{cor1}\noteref{note1}}
\author[aff1]{Zhi Wei Liu}
\author[aff1]{Ruo Yu Xing}

\affil[aff1]{School of Nuclear Science and Technology, Lanzhou University, Lanzhou 730000, China}
\corresp[cor1]{Corresponding author and speaker: sunby@lzu.edu.cn}
\authornote[note1]{To appear in the AIP Conference Proceedings of the Xiamen-CUSTIPEN Workshop on the EOS of Dense Neutron-Rich Matter in the Era of Gravitational Wave Astronomy (January 3 to 7, 2019; Xiamen China).}

\maketitle

\begin{abstract}
Comparing with a wide range of covariant energy density functional models based on the finite-range meson-exchange representation, the relativistic mean-field models with the zero-range contact interaction, namely the relativistic point-coupling models, are still infrequent to be utilized in establishing nuclear equation of state (EoS) and investigating neutron star properties, although comprehensive applications and achievements of them in describing many nuclear properties both in ground and exited states are mature. In this work, the EoS of neutron star matter is established constructively in the framework of the relativistic point-coupling models to study neutron star physics. Taking two selected functionals DD-PC1 and PC-PK1 as examples, nuclear symmetry energies and several neutron star properties including proton fractions, mass-radius relations, the core-crust transition density, the fraction of crustal moment of inertia and dimensionless tidal deformabilities are discussed. A suppression of pressure of neutron star matter found in the functional PC-PK1 at high densities results in the difficulty of its prediction when approaching to the maximum mass of neutron stars. In addition, the divergences between two selected functionals in describing neutron star quantities mentioned above are still large, ascribing to the less constrained behavior of these functionals at high densities. Then it is expected that the constraints on the dense matter EoS from precise and massive modern astronomical observations, such as the tidal-deformabilities taken from gravitational-wave events, would be essential to improve the parameterizing of the relativistic point-coupling models.
\end{abstract}

\section{Introduction}
Multi-messenger observations of a binary neutron star merger, historically first convinced by the gravitational wave event GW170817 \cite{Abbott2017PRL119.161101} together with its associated electromagnetic counterpart \cite{Abbott2017ApJL}, offer a new and unique probe for studying the nature of dense matter. From the observational data, several works have been carried out to infer constraints on a set of bulk neutron star properties such as the maximum mass, radii, and tidal deformabilities \cite{Margalit2017, Rezzolla2018, Bauswein2017, PhysRevLett.120.172703}, and subsequently constraint on the dense matter equation of state (EoS) \cite{PhysRevLett.120.172703, PhysRevLett.121.161101, PhysRevC.98.035804}. It is suggested that the parameter space of the EoS of dense matter could be sufficiently reduced by combined constraints from both terrestrial nuclear experiments and observations of neutron stars \cite{Zhang2018}. A variety of realistic EoS models, although reproducing accurately the bulk properties of finite nuclei and supporting neutron stars of two solar masses as well, are ruled out as a dense matter EoS with extrapolation after considering the tidal-deformability constraint from the gravitational-wave data.

Nuclear energy density functional (EDF) theory \cite{BenderRMP}, represented in nonrelativistic \cite{FAYANS200049} or relativistic framework \cite{VRETENAR2005101, MENG2006470}, has played an essential role in the self-consistent description of nuclei. With a few parameters, the achievements of EDF theory have been realized all over the nuclide chart. Particularly for the relativistic ones, a number of attractive features are illustrated, such as a natural treatment of the spin-orbit coupling, the origin of the empirical existence of approximate pseudospin symmetry in the nuclear single-particle spectra \cite{GINOCCHIO2005165, LIANG20151}, and a natural saturation mechanism of nuclear matter resulting from competition of various nucleon self-energies.

In recent decades, significant progresses in developing the covariant energy density functional (CDF) approaches have been made with the improved numerical techniques. Among them, the most developed and widely used version is the relativistic mean-field (RMF) approach based on the finite-range meson-exchange representation, according to the difference in treating medium dependence of effective mean-field interactions which introduces nonlinear RMF (NL-RMF) \cite{BOGUTA1983289} or density dependent RMF (DD-RMF) \cite{TYPEL1999331} models. Moreover, the CDF approach with Fock (exchange) terms was also developed in terms of the density dependent meson-nucleon coupling, namely the density dependent relativistic Hartree-Fock (DD-RHF) theory \cite{Long2006PLB640.150, Long2007PRC76.034314}, by achieving self-consistent treatments of both the nuclear tensor and spin-orbit interactions \cite{Jiang2015PRC91.034326, Zong2018CPC42.24101, Wang2018PRC98.034313}. In addition to excellent descriptions of finite nuclei, these CDF models have been widely used for studying the neutron star properties and the relevant quantities of nuclear matter such as symmetry energy and superfluidity \cite{Vretenar2003PRC68.024310, Ban2004PRC69.045805, Chen2007PRC76.054316, Sun2008PRC78.065805, BySun2010, Long2012PRC85.025806, Jiang2015PRC91.025802, Zhao2015JPG42.095101, Liu2018PRC97.025801}.

Evoking by the ease of use to study the effects beyond the mean field for nuclear low-lying collective excited states, as an alternative version, the RMF models with point-coupling interaction (PC-RMF) were proposed in which the zero-range contact interaction is used instead of the meson exchange, which is also called the relativistic point-coupling models. The medium effects can be taken into account by including higher-order (nonlinear coupling) interaction terms \cite{Nikolaus1992PRC46.1757, Burvenich2002PRC65.044308, Zhao2010PRC82.054319} or by assuming a density dependence of strength parameters for the coupling interactions \cite{Finelli2006NPA770.1, Niksic2008PRC78.034318}. Recently, the density functional PC-PK1 has been proposed by fitting to observables of 60 selected spherical nuclei, including the binding energies, charge radii, and empirical pairing gaps \cite{Zhao2010PRC82.054319}. This density functional particularly improves the description for isospin dependence of binding energies and is known to be among the most accurate density functional for the global description of nuclear masses \cite{Zhao2018IJMPE27.1830007}. As a result, it has been widely used in describing many nuclear properties, such as the Coulomb displacement energies between mirror nuclei \cite{Sun2011SCPMA54.210}, fission barriers \cite{Lu2012PRC85.011301}, nuclear quadrupole moments~\cite{Zhao2014PRC89.011301}, nuclear low-lying spectrum~\cite{Li2012PLB717.470}, nuclear magnetic and antimagnetic rotations \cite{Meng2013FP8.55, Zhao2011PLB699.181, Zhao2011PRL107.122501}, and nuclear multiple chirality~\cite{Zhao2017PLB773.1}, etc.

In the past, a wide range of CDF models have been used to establish nuclear equation of state and investigate neutron star properties. Under the finite-range meson-exchange representation, most of them adopt the mean-field approximation of meson fields, and a few of them take into account extra contributions from the exchange terms \cite{Sun2008PRC78.065805, Long2012PRC85.025806, Jiang2015PRC91.025802, Liu2018PRC97.025801, Miyatsu2012PLB709.242, Miyatsu2015AJ813.135, PhysRevC.94.045803, Li2018}. However, it is still infrequent in literature that the PC-RMF model which adopts the zero-range contact interaction is utilized maturely to investigate neutron star physics, as compared to its comprehensive application in describing the structure of finite nuclei. The possible reason may comes from the uncertainty of its EoS at high densities. Thus, in this work as a constructive study, the EoS behavior of neutron star matter arising from the PC-RMF model and the applicability of several selected effective interactions (functionals) to neutron star physics will be discussed.

\section{The RMF model with point-coupling interaction for nuclear matter}

In this section, two kinds of commonly-used PC-RMF model, namely with the nonlinear coupling terms and with a density dependence of strength parameters, will be briefly recalled for nuclear matter, which is then utilized to study various properties of neutron stars. For more details of the formulism, we refer the reader to Refs. \cite{Chen2007PRC76.054316, Zhao2010PRC82.054319, Niksic2008PRC78.034318}. The PC-RMF model is defined by a Lagrangian density that consists of only nucleon fields $\psi$. For nonlinear PC-RMF versions, it is denoted as
\begin{eqnarray}\label{eq:Lagrangian_NL}
  \mathcal{L}_{\rm{NLPC}}&=&\mathcal{L}_{\rm{free}}+\mathcal{L}_{\rm{4f}}+\mathcal{L}_{\rm{hot}}+\mathcal{L}_{\rm{der}},\\
  \mathcal{L}_{\rm{free}}&=&\bar{\psi}(\rm{i}\gamma_\mu\partial^\mu-M)\psi,\\
  \mathcal{L}_{\rm{4f}}&=&-\frac{1}{2}\alpha_S(\bar{\psi}\psi)(\bar{\psi}\psi)-\frac{1}{2}\alpha_V(\bar{\psi}\gamma_\mu\psi)(\bar{\psi}\gamma^\mu\psi)
                        -\frac{1}{2}\alpha_{TS}(\bar{\psi}\vec{\tau}\psi)(\bar{\psi}\vec{\tau}\psi)-\frac{1}{2}\alpha_{TV}(\bar{\psi}\vec{\tau}\gamma_\mu\psi)(\bar{\psi}\vec{\tau}\gamma^\mu\psi),\\
  \mathcal{L}_{\rm{hot}}&=&-\frac{1}{3}\beta_S(\bar{\psi}\psi)^3-\frac{1}{4}\gamma_S(\bar{\psi}\psi)^4
                         -\frac{1}{4}\gamma_V[(\bar{\psi}\gamma_\mu\psi)(\bar{\psi}\gamma^\mu\psi)]^2~-\frac{1}{4}\gamma_{TV}[(\bar{\psi}\vec{\tau}\gamma_\mu\psi)(\bar{\psi}\vec{\tau}\gamma^\mu\psi)]^2,\\
  \mathcal{L}_{\rm{der}}&=&-\frac{1}{2}\delta_S(\partial_\nu\bar{\psi}\psi)(\partial^\nu\bar{\psi}\psi)-\frac{1}{2}\delta_V(\partial_\nu\bar{\psi}\gamma_\mu\psi)(\partial^\nu\bar{\psi}\gamma^\mu\psi)\nonumber\\
                         &&-\frac{1}{2}\delta_{TS}(\partial_\nu\bar{\psi}\vec{\tau}\psi)(\partial^\nu\bar{\psi}\vec{\tau}\psi)-\frac{1}{2}\delta_{TV}(\partial_\nu\bar{\psi}\vec{\tau}\gamma_\mu\psi)(\partial^\nu\bar{\psi}\vec{\tau}\gamma^\mu\psi).
\end{eqnarray}
In the above, $\mathcal{L}_{\rm{free}}$ represents the kinetic term of nucleons and $\mathcal{L}_{\rm{4f}}$ is the four-fermion interactions. The higher-order terms involving more than four fermions are introduced in $\mathcal{L}_{\rm{hot}}$, reflecting the effects of medium dependence. $\mathcal{L}_{\rm{der}}$ describes derivatives in the nucleon fields, which has no contribution in the translationally invariant infinite nuclear matter. The subscripts for the coupling constants in the Lagrangian density, namely, $\alpha_S$, $\alpha_V$, $\alpha_{TS}$, $\alpha_{TV}$, $\beta_S$, $\gamma_S$, $\gamma_V$, $\gamma_{TV}$, $\delta_S$, $\delta_V$, $\delta_{TS}$, and $\delta_{TV}$, denote the tensor structure of a coupling with $S$, $V$, and $T$ standing for scalar, vector, and isovector, respectively \cite{Nikolaus1992PRC46.1757, Burvenich2002PRC65.044308, Zhao2010PRC82.054319}. For nuclear matter, the energy density $\varepsilon$ and the pressure $P$ derived from the energy-momentum tensor in the nonlinear point-coupling RMF model are given by
\begin{eqnarray}
  \varepsilon&=&\varepsilon_{\rm{kin}}^{~n}+\varepsilon_{\rm{kin}}^{~p}-\frac{1}{2}\alpha_S\rho_s^2-\frac{1}{2}\alpha_{TS}\rho_{s3}^2+\frac{1}{2}\alpha_V\rho^2+\frac{1}{2}\alpha_{TV}\rho_3^2-\frac{2}{3}\beta_S\rho_s^3-\frac{3}{4}\gamma_S\rho_s^4+\frac{1}{4}\gamma_V\rho^4+\frac{1}{4}\gamma_{TV}\rho_3^4,\label{eq:E_NL}\\
  P&=&P_{\rm{kin}}^{~n}+P_{\rm{kin}}^{~p}+\frac{1}{2}\alpha_S\rho_s^2+\frac{1}{2}\alpha_{TS}\rho_{s3}^2+\frac{1}{2}\alpha_V\rho^2+\frac{1}{2}\alpha_{TV}\rho_3^2+\frac{2}{3}\beta_S\rho_s^3+\frac{3}{4}\gamma_S\rho_s^4+\frac{3}{4}\gamma_V\rho^4+\frac{3}{4}\gamma_{TV}\rho_3^4.\label{eq:P_NL}
\end{eqnarray}
In practical application of the nonlinear point-coupling model, the widely used nonlinear PC-RMF functionals include PC-LA \cite{Nikolaus1992PRC46.1757} and PC-F1 \cite{Burvenich2002PRC65.044308}. Recently, a new parameter sets PC-PK1 \cite{Zhao2010PRC82.054319} is proposed. In particular, PC-PK1 provides a good description for the isospin dependence of binding energy along either the isotopic or the isotonic chain, which makes it reliable for application in exotic nuclei. In this work, we just take the functional PC-PK1 as an example to study neutron star physics within the nonlinear PC-RMF model.

As an alternative version, the density-dependent PC-RMF model introduce a density dependence in the strength parameters of coupling interactions. Its Lagrangian density can be written as
\begin{eqnarray}\label{eq:Lagrangian_DD}
  \mathcal{L}_{\rm{DDPC}}&=&\mathcal{L}_{\rm{free}}+\mathcal{L}_{\rm{4f}}+\mathcal{L}_{\rm{der}},\\
  \mathcal{L}_{\rm{free}}&=&\bar{\psi}(\rm{i}\gamma_\mu\partial^\mu-M)\psi,\\
  \mathcal{L}_{\rm{4f}}&=&-\frac{1}{2}G_S(\rho)(\bar{\psi}\psi)(\bar{\psi}\psi)-\frac{1}{2}G_V(\rho)(\bar{\psi}\gamma_\mu\psi)(\bar{\psi}\gamma^\mu\psi)\nonumber\\
                        &&-\frac{1}{2}G_{TS}(\rho)(\bar{\psi}\vec{\tau}\psi)(\bar{\psi}\vec{\tau}\psi)-\frac{1}{2}G_{TV}(\rho)(\bar{\psi}\vec{\tau}\gamma_\mu\psi)(\bar{\psi}\vec{\tau}\gamma^\mu\psi),\\
  \mathcal{L}_{\rm{der}}&=&-\frac{1}{2}D_S(\rho)(\partial_\nu\bar{\psi}\psi)(\partial^\nu\bar{\psi}\psi).
\end{eqnarray}
In the above, $\mathcal{L}_{\rm{free}}$ denotes the kinetic term of the nucleons, $\mathcal{L}_{\rm{4f}}$ is the four-fermion interactions, and $\mathcal{L}_{\rm{der}}$ represents derivatives in the nucleon scalar densities. Unlike the nonlinear point-coupling model, the density-dependent one used here includes only second-order interaction terms with density-dependent couplings $G_i(\rho)$ and $D_i(\rho)$ that are determined from finite-density QCD sum rules and in-medium chiral perturbation theory \cite{Finelli2006NPA770.1}. Following the standard procedure, the energy density $\varepsilon$ and the pressure $P$ for nuclear matter are then derived from the energy-momentum tensor and represented as
\begin{eqnarray}
  \varepsilon&=&\varepsilon_{\rm{kin}}^{~n}+\varepsilon_{\rm{kin}}^{~p}-\frac{1}{2}G_S\rho_s^2-\frac{1}{2}G_{TS}\rho_{s3}^2+\frac{1}{2}G_V\rho^2+\frac{1}{2}G_{TV}\rho_3^2,\label{eq:E_DD}\\
  P&=&P_{\rm{kin}}^{~n}+P_{\rm{kin}}^{~p}+\frac{1}{2}G_S\rho_s^2+\frac{1}{2}G_{TS}\rho_{s3}^2+\frac{1}{2}G_V\rho^2+\frac{1}{2}G_{TV}\rho_3^2\nonumber\\
    &&+\frac{1}{2}\frac{\partial G_S}{\partial\rho}\rho_s^2\rho+\frac{1}{2}\frac{\partial G_{TS}}{\partial\rho}\rho_{s3}^2\rho+\frac{1}{2}\frac{\partial G_V}{\partial\rho}\rho^3+\frac{1}{2}\frac{\partial G_{TV}}{\partial\rho}\rho_3^2\rho.\label{eq:P_DD}
\end{eqnarray}
In this work, we just select DD-PC1 \cite{Niksic2008PRC78.034318} as the functional of the density-dependent PC-RMF model, which was proposed from the EoS of nuclear matter and the masses of 64 axially deformed nuclei in the mass regions $A\simeq150-180$ and $A\simeq230-250$.

To study asymmetric nuclear matter and neutron star matter, it is helpful to introduce several quantities which can describe the dependence of the equation of state on density or isospin asymmetry. The EoS of asymmetric nuclear matter at zero temperature is defined by its binding energy per nucleon $E(\rho,\delta)$, where $\rho=\rho_n+\rho_p$ denotes the baryon density, and $\delta\equiv(\rho_n-\rho_p)/(\rho_n+\rho_p)$ is the isospin asymmetry with $\rho_{n/p}$ the neutron/proton density. Conventionally, due to the difficulty of analytical extraction, the various order of nuclear symmetry energies can be approximately extracted by expanding the zero-temperature EoS in a Taylor series with respect to the $\delta$. Within this approximation, the EoS is then expressed as
\begin{equation}
    E(\rho,\delta)
    ~=~E_0(\rho)~+~S_{2}(\rho)\delta^{2}~+~S_{4}(\rho)\delta^{4}+\cdots,\label{eq:Taylor expand}
\end{equation}
where $E_0(\rho)=E(\rho,\delta=0)$ denotes the EoS of symmetric nuclear matter, and the coefficients $S_2(\rho)$ and $S_4(\rho)$ give the density-dependent second-order and fourth-order symmetry energy, respectively. Here the odd-order terms of the expansion are discarded due to the assumption of the charge-independence of nuclear force and the neglecting of the Coulomb interaction in infinite nuclear matter. The density slope parameter $L$ is used to reflect the density dependence of $S_2(\rho)$ at saturation density $\rho_0$, which is defined as
\begin{equation}
\label{eq:slope parameter}
  L=3\rho_0\frac{\partial S_2(\rho)}{\partial\rho}\bigg|_{\rho=\rho_0}.
\end{equation}

\begin{figure}[h]
  \centerline{\includegraphics[width=220pt]{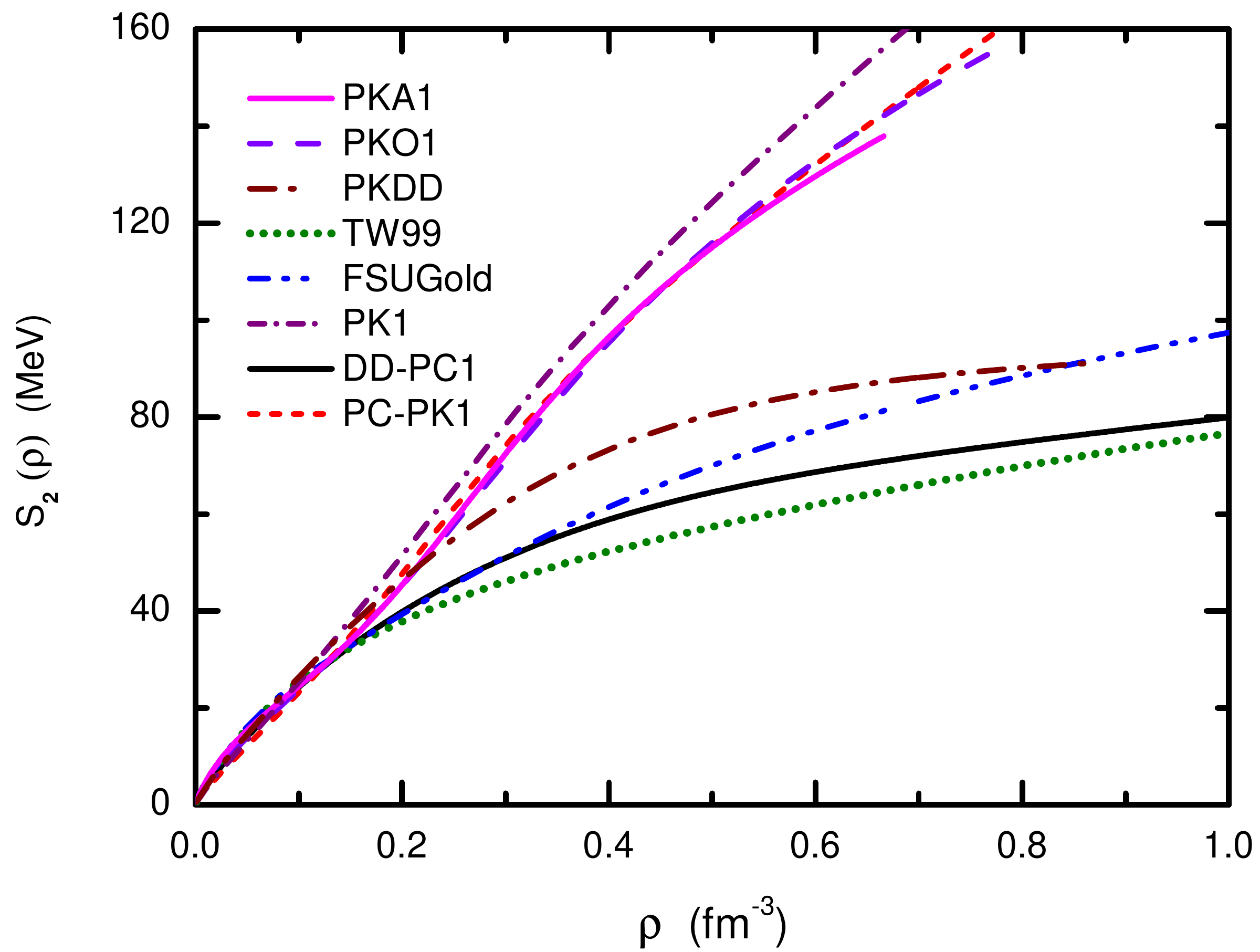}}
  \caption{The nuclear symmetry energy $S_2(\rho)$ as a function of the baryon density $\rho$. The results are given by PC-RMF functionals DD-PC1 and PC-PK1, in comparison with those within the finite-range meson-exchange CDF models.}\label{Fig:symmetry_energy}
\end{figure}

In Fig. \ref{Fig:symmetry_energy} the baryon density dependence of nuclear symmetry energy $S_2(\rho)$ is plotted with two PC-RMF functionals DD-PC1 \cite{Niksic2008PRC78.034318} and PC-PK1 \cite{Zhao2010PRC82.054319}. For a brief comparison, the results from six featured CDF functionals with meson-exchange representation are also calculated, namely, the DD-RHF ones PKA1 \cite{Long2007PRC76.034314} and PKO1 \cite{Long2006PLB640.150}, the DD-RMF ones PKDD \cite{Long2004PRC69.034319} and TW99 \cite{Typel1999NPA656.331}, and the NL-RMF ones FSUGold \cite{Todd2005PRL95.122501} and PK1 \cite{Long2004PRC69.034319}. Two PC-RMF functionals DD-PC1 and PC-PK1 generate clear difference of $S_2(\rho)$ at suprasaturation densities. The divergence can be found at saturation density as well, as seen in Table \ref{tab:bulk}, which is mainly associated with the difference of the potential part $S_{2,\rm{pot}}$. For the definition of the kinetic contribution to the energy densities $\varepsilon_{\rm{kin}}$ in PC-RMF models, which contributes to $S_{2,\rm{kin}}$ and $S_{4,\rm{kin}}$, we follow that in Ref. \cite{Chen2007PRC76.054316}. It should be noticed that such a definition could be different in the CDF approaches with meson-exchange representation, e.g. see Ref. \cite{Sun2008PRC78.065805}.

\begin{table}[h]
  \caption{Bulk properties of symmetric nuclear matter at saturation density $\rho_0$ (in units of $\rm{fm}^{-3}$), i.e., the symmetry energy $S_2(\rho_0)$, and the fourth-order symmetry energy $S_4(\rho_0)$ (in units of MeV). $S_{2,\rm{pot}}$ (or $S_{4,\rm{pot}}$) and $S_{2,\rm{kin}}$ (or $S_{4,\rm{kin}}$) correspond to the potential part and the kinetic part of $S_2(\rho_0)$ (or $S_4(\rho_0)$), respectively. The results are calculated by using the PC-RMF functionals DD-PC1 and PC-PK1.}
  \label{tab:bulk}
  \tabcolsep12pt
  \begin{tabular}{lrcccccc}
    \hline
     &\tch{1}{c}{b}{$\rho_0$} &\tch{1}{c}{b}{$S_2(\rho_0)$} &\tch{1}{c}{b}{$S_{2,\rm{pot}}$} &\tch{1}{c}{b}{$S_{2,\rm{kin}}$} &\tch{1}{c}{b}{$S_4(\rho_0)$} &\tch{1}{c}{b}{$S_{4,\rm{pot}}$} &\tch{1}{c}{b}{$S_{4,\rm{kin}}$} \\
    \hline
    DD-PC1      & 0.152     & 33.01     & 4.44    & 28.57     &0.65     &-0.17     &0.82 \\
    PC-PK1      & 0.153     & 35.61     & 7.78    & 27.83     &0.65     &-0.14     &0.79 \\
    \hline
  \end{tabular}
\end{table}

\section{Neutron star properties within PC-RMF models}
\begin{figure}[h]
  \centerline{\includegraphics[width=220pt]{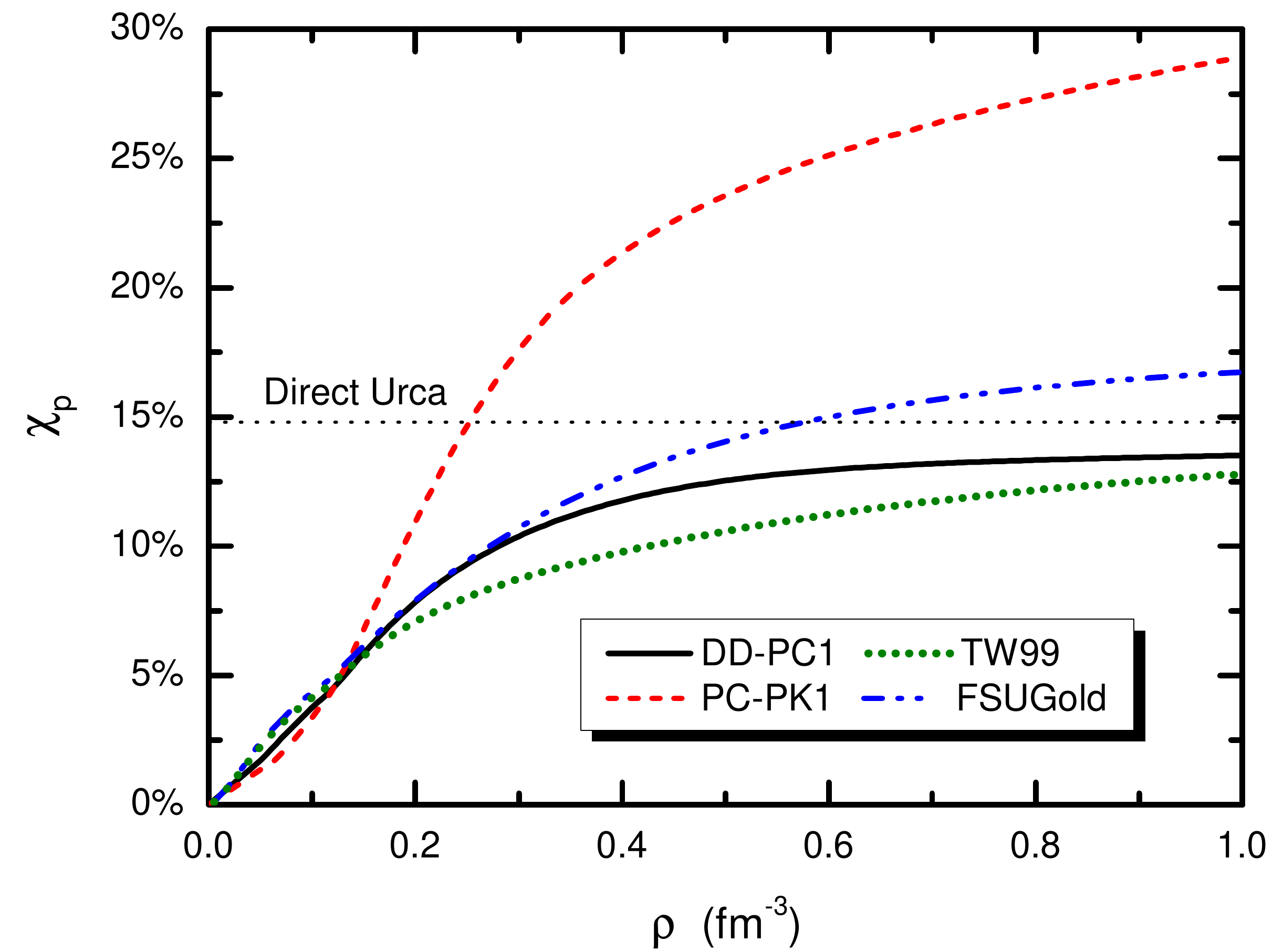}}
  \caption{Proton fractions $\chi_p=\rho_p/(\rho_n+\rho_p)$ in neutron star matter. The results are calculated by PC-RMF functionals DD-PC1 and PC-PK1, in comparison with those using DD-RMF functional TW99 and NL-RMF one FSUGold. The dotted line ($\chi_p=14.8\%$) represents the threshold for the occurrence of the DUrca process by assuming muons in the massless limit.}\label{Fig:proton_fraction}
\end{figure}

In this work a neutron star will be represented as the $\beta$-stable nuclear matter system, which consists of nucleons (neutrons and protons) and leptons $\lambda$ (mainly $e^-$ and $\mu^-$). The inclusion of extra degrees of freedom beyond nucleons, e.g., hyperons, mesons, and quarks, will not be discussed. With the inclusion of leptons, one should impose the $\beta$-equilibrium, baryon density conservation, and charge neutrality conditions so as to obtain the nucleon(lepton) fractions and then the EoS of neutron star matter for a given baryon density and isospin asymmetry. From these restrictions, it has been proved \cite{Cai2012PRC85.024302, Liu2018PRC97.025801} that the proton fraction $\chi_p\equiv\rho_p/(\rho_n+\rho_p)$ can be expressed as
\begin{eqnarray}\label{eq:neutron_star_matter}
  \chi_p(\rho)=\frac{1}{3\pi^2\rho}\sum_\lambda\left\{\left[2\frac{\partial E(\rho,\delta)}{\partial\delta}\right]^2-m_\lambda^2\right\}^{3/2},
\end{eqnarray}
which provide a convenient treatment to determine the isospin asymmetry $\delta$ and construct the EoS of neutron star matter. In Fig. \ref{Fig:proton_fraction}, the proton fractions $\chi_p$ in neutron star matter given by the selected CDF functionals are plotted as a
function of baryon density. Within PC-RMF calculations, a stronger density dependence of the proton fraction $\chi_p$ is obtained by PC-PK1 than DD-PC1, which can be well associated with the behavior of the symmetry energy shown in Fig. \ref{Fig:symmetry_energy}. As has been illustrated in many literatures, e.g. in Ref. \cite{Sun2008PRC78.065805}, the stronger the density dependence of symmetry energy is, the more difficult it becomes for the system to become asymmetric and the easier it is for neutrons to decay into protons and electrons, which results in smaller neutron abundance and larger proton, electron, and muon abundances in neutron stars.

The model dependence of the proton fraction at high densities affects theoretical prediction of the cooling mechanism of neutron stars sensitively, among which the occurrence of the DUrca process is mostly discussed \cite{BOGUTA1981255, Lattimer1991PRL66.2701}. The cooling rate of neutron stars could be enhanced significantly via the DUrca process, i.e. $n\rightarrow p+e^-+\bar{\nu}_e$ and $p+e^-\rightarrow n+\nu_e$, leading the star to cool off rapidly by emitting the thermal neutrinos. From modern observational soft X-ray data of cooling neutron stars associated with popular synthesis model analyses, it is suggested that an acceptable EoS would not allow the DUrca process to occur in canonical neutron stars with 1.4~$M_\odot$ \cite{Klahn2006}, indicating the strict constraints on the proton fraction of neutron star matter at high densities. By assuming muons in the massless limit, the threshold for the occurrence of the DUrca process is denoted as $\chi_p=14.8\%$. Thus, it is extracted from Fig. \ref{Fig:proton_fraction} that PC-PK1 predicts a explicitly small baryon density $\rho^{DU}\simeq0.25~{\rm fm^{-3}}$ for the DUrca process occurring in the center of neutron stars, which corresponds to a fairly low star mass 1.06~$M_\odot$. For the NL-RMF functional FSUGold, the DUrca process will occur when the mass is larger than 1.49~$M_\odot$ and central density $\rho\gtrsim0.58~{\rm fm^{-3}}$, while the values of $\chi_p$ given by TW99 and DD-PC1 do not support the onset of DUrca process at all.

\begin{figure}[h]
  \centerline{\includegraphics[width=220pt]{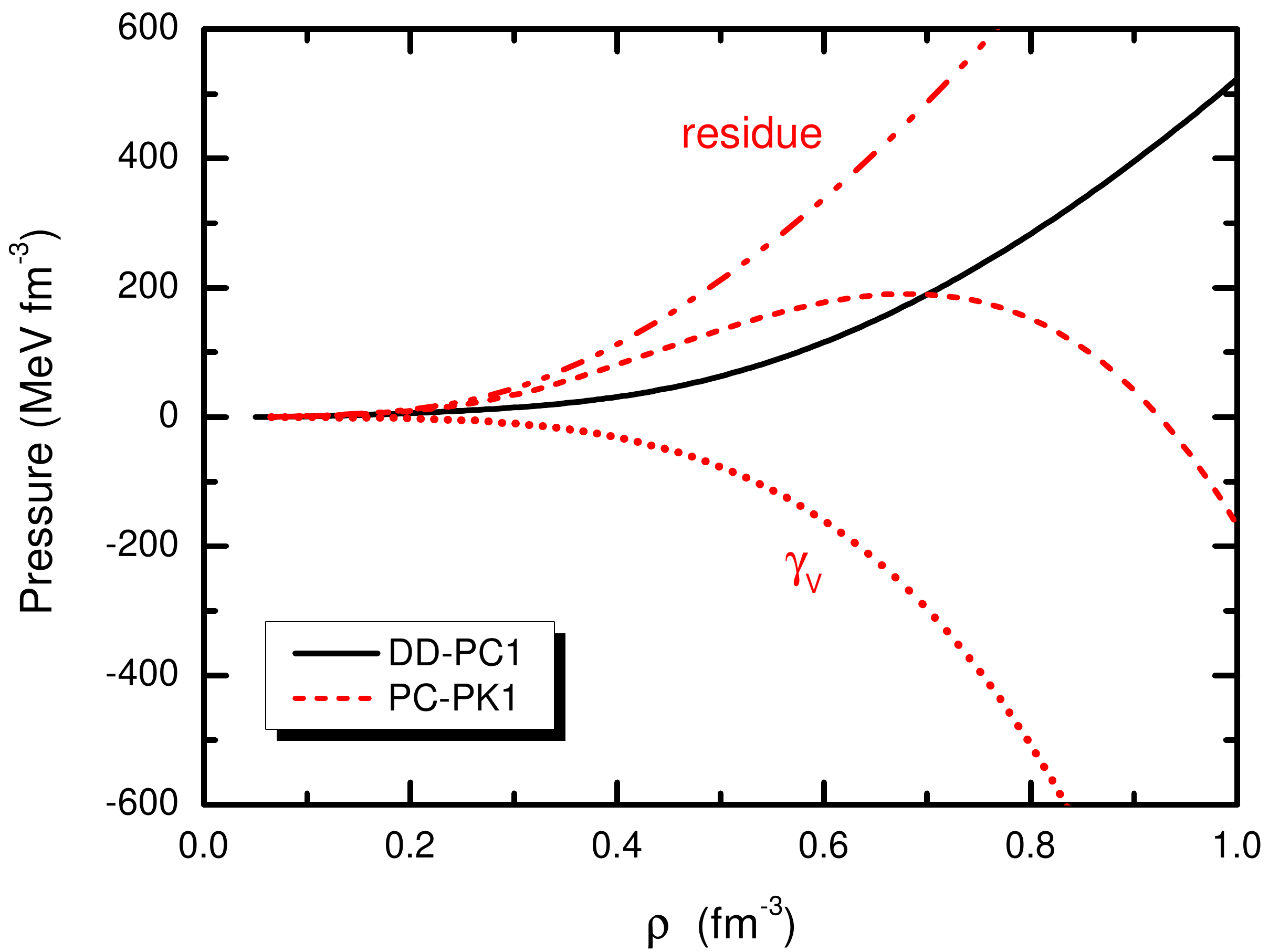}}
  \caption{The pressure of neutron star matter as a function of the baryon density $\rho$. The results are calculated by PC-RMF functionals DD-PC1 (solid line) and PC-PK1 (dashed line). For PC-PK1, the contribution from $\gamma_V$ relevant term and the residual ones are plotted as well, see Eq. \ref{eq:P_NL} for details.}\label{Fig:pressure}
\end{figure}

The variation of the pressure with respect to density is crucial to understand the structure of neutron stars. A stronger density dependence of the pressure at high densities would lead to a larger value of the maximum mass for neutron stars that can be sustained against collapse. Figure \ref{Fig:pressure} shows the pressures of neutron star matter calculated by PC-RMF functionals as a function of the baryon density. DD-PC1 shows a trend that the pressure increases monotonically with the density, while the pressure given by PC-PK1 rises first at low densities and then drops down after reaching the maximum value at about $4\thicksim5~\rho_0$. A negative pressure at high densities may lead to problem when looking for maximum mass limits of neutron stars, since the allowed EoS should be consistent with the observational constraint provided by the existence of 2~$M_\odot$ neutron stars \cite{Antoniadis1233232}. After checking the contribution from each term given in Eq. \ref{eq:P_NL}, it is seen for PC-PK1 the suppression of pressure is mainly ascribed to $\gamma_V$ relevant term, which is always negative and goes down with increasing density. The similar case is found in nonlinear PC-RMF functionals PC-LA \cite{Nikolaus1992PRC46.1757} and PC-F1 \cite{Burvenich2002PRC65.044308} as well.

\begin{figure}[h]
  \centerline{\includegraphics[width=220pt]{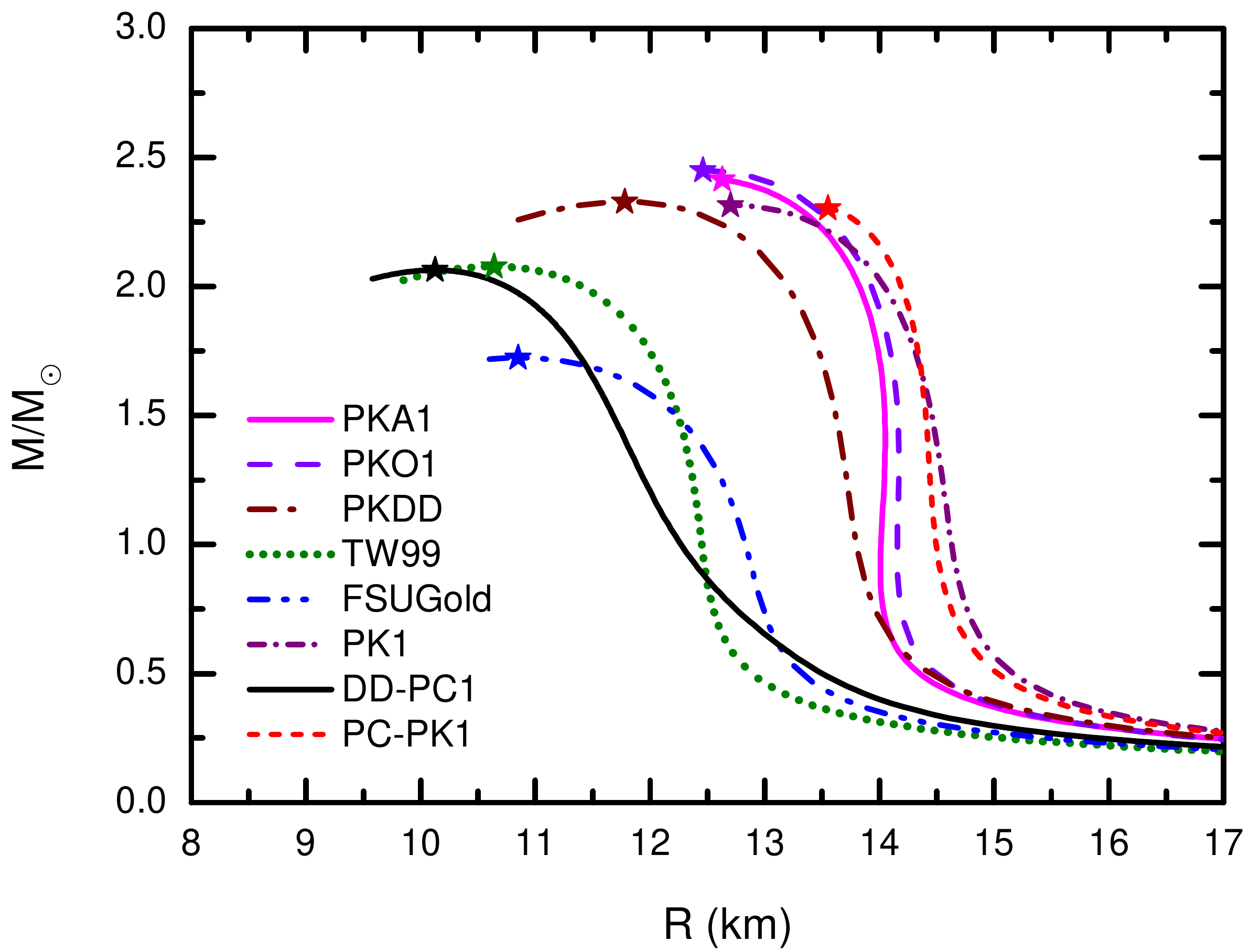}}
  \caption{Mass-radius relations of neutron stars calculated by the selected CDF EoSs. The corresponding maximum masses are marked by filled star symbols.}\label{Fig:M_R}
\end{figure}

To investigate the structure of neutron stars, it is also important to include the physics of crust region. In this work, the EoSs of neutron star matter under $\beta$-equilibrium given above are used at high densities (neutron star core region), in combination with the BPS \cite{BPS} and BBP \cite{BBP} EoS at low densities (neutron star crust region). The EoS of neutron star core and crust is matched by the core-crust transition pressure, which will be discussed later in the next section. The structure of a static, spherically symmetric neutron star is then obtained by solving the stellar hydrostatic equilibrium equations, namely the Tolman-Oppenheimer-Volkov (TOV) equations \cite{Oppenheimer1939PR55.374, Tolman1939PR55.364}. By taking $c=G=1$, the TOV equations become
\begin{eqnarray}\label{eq:TOV}
  \frac{dP}{dr}&=&-\frac{[P(r)+\varepsilon(r)][M(r)+4\pi r^3P(r)]}{r[r-2M(r)]},\\
  \frac{dM}{dr}&=&4\pi r^2\varepsilon(r),
\end{eqnarray}
where $P(r)$ is the pressure of the star at radius $r$, and $M(r)$ defines the total star mass inside a sphere of radius $r$. For a given
EoS, the TOV equation has the unique solution that depends on a single parameter characterizing the conditions of matter at the center, such as the central density $\rho(0)$ or the central pressure $P(0)$. Using the EoSs of the selected CDF functionals, the masses and radii of neutron stars are drawn in Fig. \ref{Fig:M_R}. Because of the suppressed behavior of pressure at high densities, the mass-radius curve for PC-RMF functional PC-PK1 stops at the mass around 2.30~$M_\odot$ and can not reach the maximum mass limits. Nevertheless, the existence of 2~$M_\odot$ neutron stars is permitted by these CDF EoSs except NL-RMF one FSUGold. In addition, it is revealed that PC-RMF functional DD-PC1 predicts smaller radius for a canonical neutron star than others.

\subsection{Core-crust transition density of neutron stars}
The core-crust interface of neutron stars, with the phase transition between nuclei and uniform matter, can be decided by study the instability of matter in neutron stars \cite{Kubis2007PRC76.025801, Lattimer2007Phyreport442.109}. Both dynamical and thermodynamical methods could be used to determine the stability of the uniform ground state against cluster formation \cite{Xu2009AJ697.1549, Fattoyev2010PRC82.025810, Ducoin2011PRC83.045810, Piekarewicz2014PRC90.015803, Seif2014PRC89.028801, Atta2014PRC90.035802, Liu2018PRC97.025801}. Here the thermodynamical method is utilized to determine the core-crust transition density $\rho_t$ which separates the liquid core from the inner crust in neutron stars. It has been compared that the dynamical method predicts a slightly smaller transition density, about $0.005\thicksim0.015~\rm{fm}^{-3}$ lower, than the thermodynamical calculation \cite{Ducoin2011PRC83.045810}.

\begin{figure}[h]
  \centerline{\includegraphics[width=300pt]{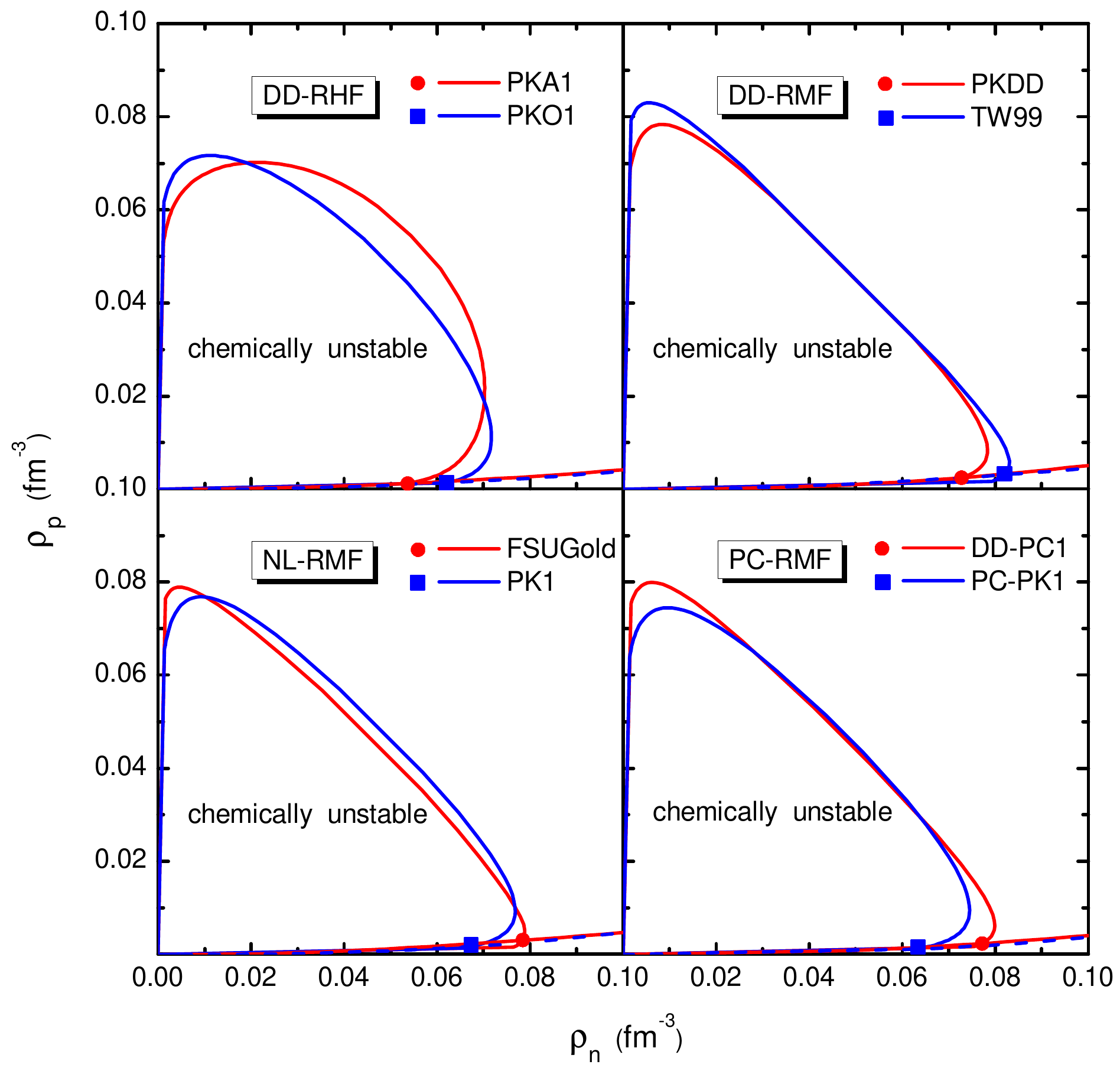}}
  \caption{Chemical instability boundaries shown in the $\rho_n\sim\rho_p$ plane using the various CDF functionals at zero temperature. $\rho_n$ vs. $\rho_p$ for neutron star matter are shown as the solid (red) and dash (blue) line. The core-crust transition density from the thermodynamical method are indicated with the filled dot (red) and square (blue).}\label{Fig:phase_diagram}
\end{figure}

In the thermodynamical method, the stability of uniform matter is required to obey the intrinsic stability condition of any single phase, namely,
\begin{equation}\label{eq:stability1}
      -~\left(\frac{\partial P}{\partial v}\right)_{\mu_{np}}~>~0, \qquad
      -~\left(\frac{\partial\mu_{np}}{\partial q_c}\right)_v~>~0.
\end{equation}
where $P$ is the total pressure of neutron star matter, $v=1/\rho$ denotes the average volume per baryon, $\mu_{np}=\mu_n-\mu_p$ represents the difference between neutron and proton chemical potentials, and $q_c$ corresponds to the average charge per baryon. Here the finite size effects due to surface and Coulomb energies of nuclei are ignored. In addition, by introducing a density dependent thermodynamical potential $V_{ther}(\rho)$, the stability condition of Eq. (\ref{eq:stability1}) can be equivalently expressed as:
\begin{eqnarray}\label{eq:stability2}
  V_{ther}(\rho)\equiv 2\rho\frac{\partial E(\rho,\delta)}{\partial\rho}
  ~+~\rho^2\frac{\partial^2E(\rho,\delta)}{\partial\rho^2}
  -~\left[\rho\frac{\partial^2E(\rho,\delta)}{\partial\rho\partial\delta}\right]^2\left/\frac{\partial^2E(\rho,\delta)}{\partial\delta^2}\right.~>~0,
\end{eqnarray}
subsequently the threshold with $V_{ther}(\rho_t)=0$ determine the critical density $\rho_t$ for the core-crust transition. To solve this equation, it is convenient to plot lines which satisfy $V_{ther}=0$ in the $\rho_n\sim\rho_p$ plane so as to separate chemically unstable and stable regions, as shown in Fig. \ref{Fig:phase_diagram}. Their intersection points to the curves $\rho_p/\rho_n$ for $\beta$-equilibrium neutron star matter then give the values of $\rho_t$ for various CDF functionals. In Table \ref{tab:core-crust}, we summarize the values of several quantities relevant for the core-crust transition. It is suggested from various theoretical models that a good anti-correlation between the transition density $\rho_t$ and the symmetry energy slope $L$ exists \cite{Xu2009AJ697.1549, Fattoyev2010PRC82.025810, Ducoin2011PRC83.045810, Providencia2014, Liu2018PRC97.025801}, which is still fulfilled by the PC-RMF functionals selected here.

\begin{table}[h]
  \caption{The core-crust transition density $\rho_t$ (in units of $\rm{fm}^{-3}$), and the corresponding values of the proton fraction $\chi_p$ and pressure $P_t$ (in units of MeV $\rm{fm}^{-3}$) at $\rho_t$ in neutron stars, as well as the density slope of symmetry energy $L$ (in units of MeV) with the various CDF functionals.}
  \label{tab:core-crust}
  \tabcolsep14.5pt
  \begin{tabular}{clcccrc}
    \hline
      &Interaction &\tch{1}{c}{b}{$\rho_t$} &\tch{1}{c}{b}{$\chi_t$} &\tch{1}{c}{b}{$P_t$} &\tch{1}{c}{b}{$L$} &$\rm{Ref.}$ \\
    \hline
    DD-RHF &PKA1        & 0.0550     & 0.0235     & 0.2567     & 103.5     & \cite{Long2007PRC76.034314} \\
           &PKO1        & 0.0634     & 0.0219     & 0.3023     & 97.7      & \cite{Long2006PLB640.150} \\
    DD-RMF &PKDD        & 0.0755     & 0.0332     & 0.6142     & 90.2      & \cite{Long2004PRC69.034319} \\
           &TW99        & 0.0851     & 0.0367     & 0.5243     & 55.3      & \cite{Typel1999NPA656.331} \\
    NL-RMF &FSUGold     & 0.0815     & 0.0373     & 0.4868     & 60.5      & \cite{Todd2005PRL95.122501} \\
           &PK1         & 0.0693     & 0.0252     & 0.5070     & 115.9     & \cite{Long2004PRC69.034319} \\
    PC-RMF &DD-PC1      & 0.0794     & 0.0294     & 0.4850     & 70.1      & \cite{Niksic2008PRC78.034318} \\
           &PC-PK1      & 0.0629     & 0.0167     & 0.3184     & 112.9     & \cite{Zhao2010PRC82.054319} \\
    \hline
  \end{tabular}
\end{table}

\section{Fraction of crustal moment of inertia of neutron stars}
The moments of inertia, reflecting mass distribution in neutron stars interior, provides a powerful probe of their internal structure. It is suggested that the measured moment of inertia, such as from spin-orbit coupling in double pulsar systems \cite{Lyne2004SCI303.1153, Kramer2009CQG26.073001}, would delimit EoS significantly \cite{Lattimer2005ApJ629979} and be used to distinguish neutron stars from quark stars \cite{Yagi2013SCI341.365}. Besides, the moment of inertia that resides in the crust of neutron stars plays an important role in understanding the mechanism of pulsar glitches. Several theoretical efforts devoted \cite{Link1999PRL83.3362,  Andersson2012PRL109.241103, ChamelPRL2013} and were compared to the glitch activities of Vela pulsar \cite{Baym1969NAT224.872}, in order to constrain the fraction of crustal moment of inertia, although uncertainties still exist such as in the calculation of entrainment of superfluid neutrons in the crust \cite{Andersson2012PRL109.241103, ChamelPRL2013, LiAAstrophysicalJournal2016, Watanabe2017PRL119.062701}. From our recent work within CDF theory, it is seen that the crustal moment of inertia could be taken as a more sensitive probe of the neutron-star matter EoS rather than the total one \cite{Qian2018}.

Based on the slowly rotating assumptions, the moment of inertia of neutron stars are defined according to the general relativity \cite{Hartel1967APJ150.1005, Lattimer2007Phyreport442.109}
\begin{eqnarray}\label{totalI}
\frac{dI}{dr}=-\frac{2c^2}{3G}r^3\omega(r)\frac{dj(r)}{dr},
\end{eqnarray}
where $j(r)=e^{-(v(r)+\lambda(r))/2}$, with the metric functions $\nu(r)$ and $\lambda(r)$ satisfying
\begin{eqnarray}
  \frac{dv(r)}{dr}&=2G\displaystyle\frac{M(r)+4{\pi}r^3p(r)/c^2}{r(r-2GM(r)/c^2)},\\
  e^{-\lambda(r)}&=1-\displaystyle\frac{2GM(r)}{rc^2}.
\end{eqnarray}
The rotational drag $\omega(r)$ can be solved from the equation
\begin{eqnarray}\label{omega}
  \frac{d}{dr}(r^4j(r)\frac{d\omega(r)}{dr})=-4r^3\omega(r)\frac{dj(r)}{dr},
\end{eqnarray}
with the boundary conditions required by the continuity at the stellar surface
\begin{eqnarray}\label{omega2}
  \omega(R)&=1-\displaystyle\frac{2GI}{R^3c^2},~~~~j(R)=1.
\end{eqnarray}
Starting from a constant trial value of $\omega$ and $d\omega/dr=0$ at $r=0$, the stellar profile of (crustal) moment of inertia can therefore be numerically obtained by solving the equations above iteratively together with the TOV equations.

\begin{figure}[h]
  \centerline{\includegraphics[width=220pt]{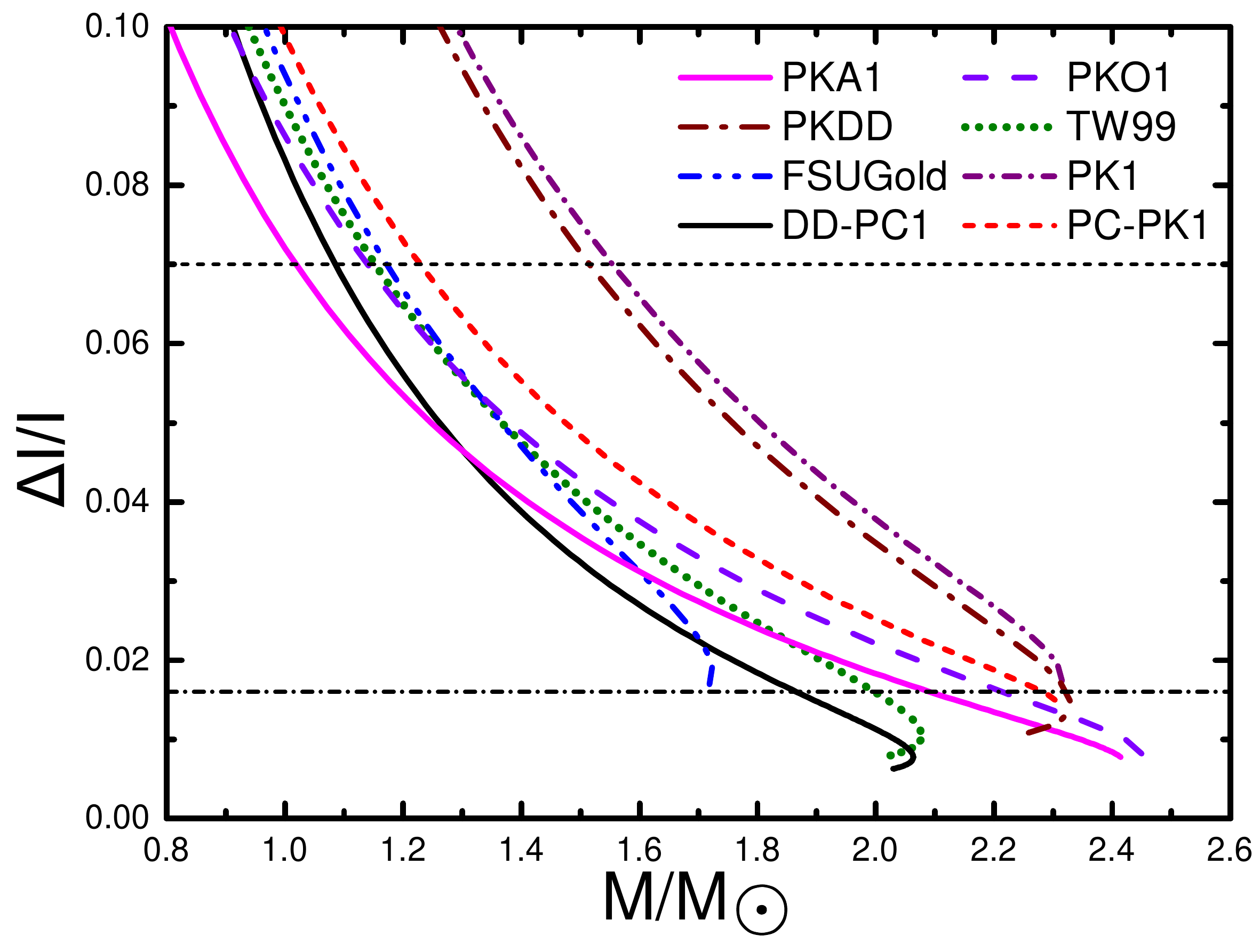}}
  \caption{The fraction of crustal moment of inertia $\Delta I/I$ as a function of the neutron star mass (in unit of the solar mass). Two horizontal lines represent the constraints on $\Delta I/I$, namely $\Delta I/I \leq$ 0.016 \cite{Andersson2012PRL109.241103} and $\Delta I/I \leq$ 0.07 \cite{Chamel2012PRC85.035801}, respectively.}\label{Fig:M-DII}
\end{figure}

In Fig. \ref{Fig:M-DII}, the stellar mass dependence of the fraction of crustal moment of inertia $\Delta I/I$ is depicted with the selected CDF functionals. It is shown in all results that $\Delta{I}/I$ decreases monotonically as the stellar mass increases until the maximum mass. For neutron stars with $M\gtrsim 1.35~M_\odot$, PC-RMF functional DD-PC1 gives the smallest values of $\Delta I/I$ among the selected CDF ones, which is mainly attributed to the relatively small star radius predicted by DD-PC1. A better comprehension could be got by using the approximated expression of the ratio, which is suggested by \cite{Lattimer2007Phyreport442.109}
\begin{equation}  \label{jiexi}
\frac{\Delta{I}}{I}\simeq\frac{8{\pi}P_tR^4}{3GM^2}\left[\frac{MR^2}{I}-2\beta\right]e^{-4.8\Delta{R}/R}.
\end{equation}
Therefore, it is seen that $\Delta{I}/{I}$ is mainly dominated by the core-crust transition pressure $P_t$ at the density $\rho_t$ and the radius $R$ of neutron star as well. Correspondingly, both the core-crust transition properties at subsaturation density and the density dependent behavior of EoS at high densities play the role in determining $\Delta{I}/{I}$.

\section{Tidal deformabilities of neutron stars}
\begin{figure}[h]
  \centerline{\includegraphics[width=220pt]{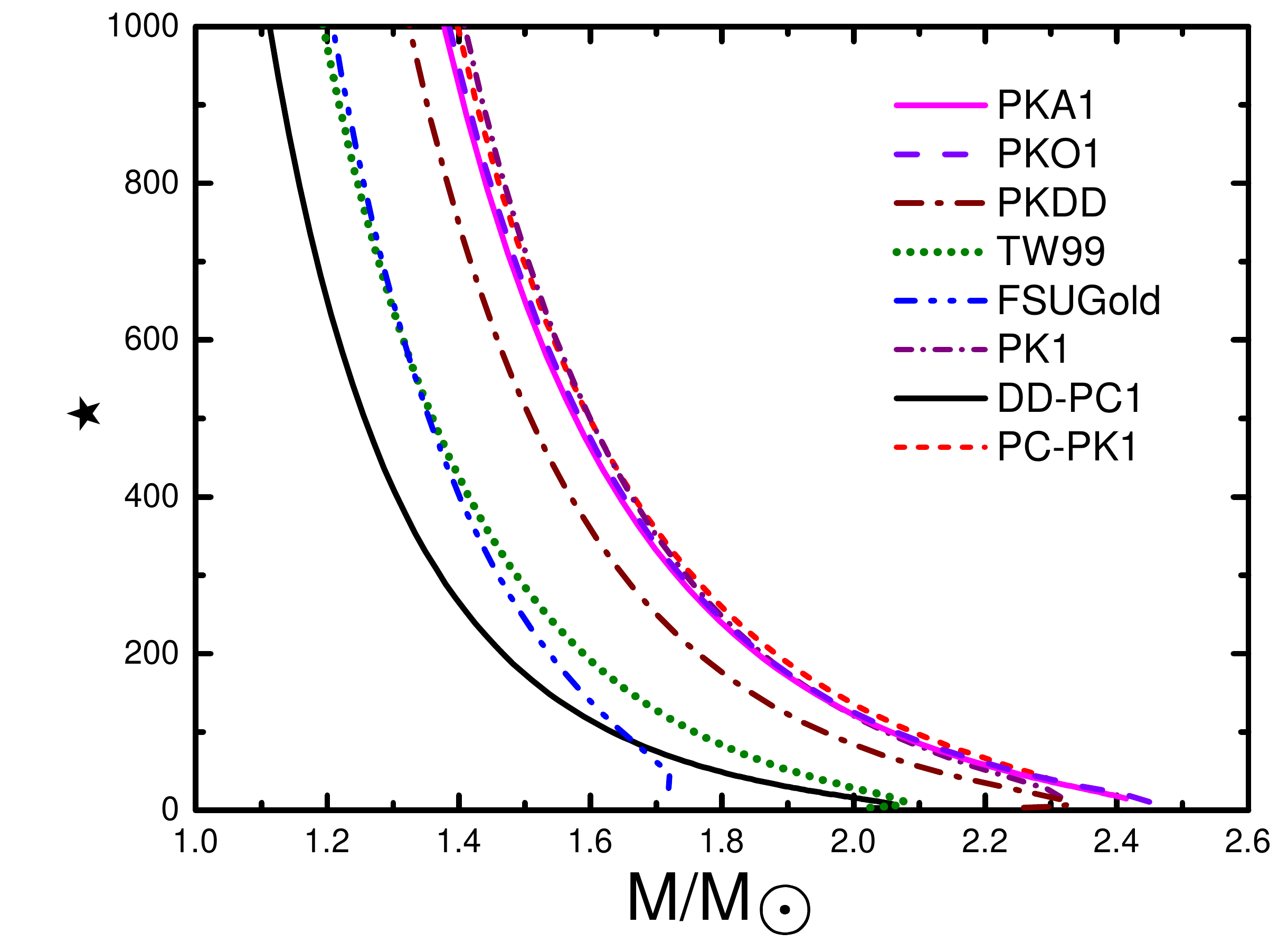}}
  \caption{The dimensionless tidal deformability $\Lambda_{\bigstar}$ of neutron stars as a function of stellar mass (in unit of the solar mass).}\label{Fig:L-M}
\end{figure}

The dimensionless tidal deformability $\Lambda_{\bigstar}$ quantifies the degree of quadrupole deformation of a neutron star due to the tidal field induced by its companion, which can be described as
\begin{eqnarray}\label{eq:lambda}
    \Lambda_{\bigstar}~=~\frac{2}{3}k_{2}\left(\frac{c^{2}R}{GM}\right)^5,
\end{eqnarray}
where $k_{2}$ is the second Love number \cite{Hinderer2008, PhysRevD.80.084035}. It is clear that the value of $\Lambda_{\bigstar}$ is essentially sensitive to the compactness parameter $GM/c^2R$, corresponding sensitive to the nature of the dense matter EoS. In Fig. \ref{Fig:L-M}, the dimensionless tidal deformability is represented as a function of the neutron star mass, using the PC-RMF functionals DD-PC1 and PC-PK1 as compared to the other CDF ones. The values of $\Lambda_{\bigstar}$ decrease systematically with increasing stellar mass, but show a sizable spread across the selected models. In a wide range of stellar mass, DD-PC1 predicts the smallest values of $\Lambda_{\bigstar}$, which are consistent with the relatively small value of stellar radii as has been discussed for Fig. \ref{Fig:M_R}.

\begin{figure}[t]
  \centerline{\includegraphics[width=220pt]{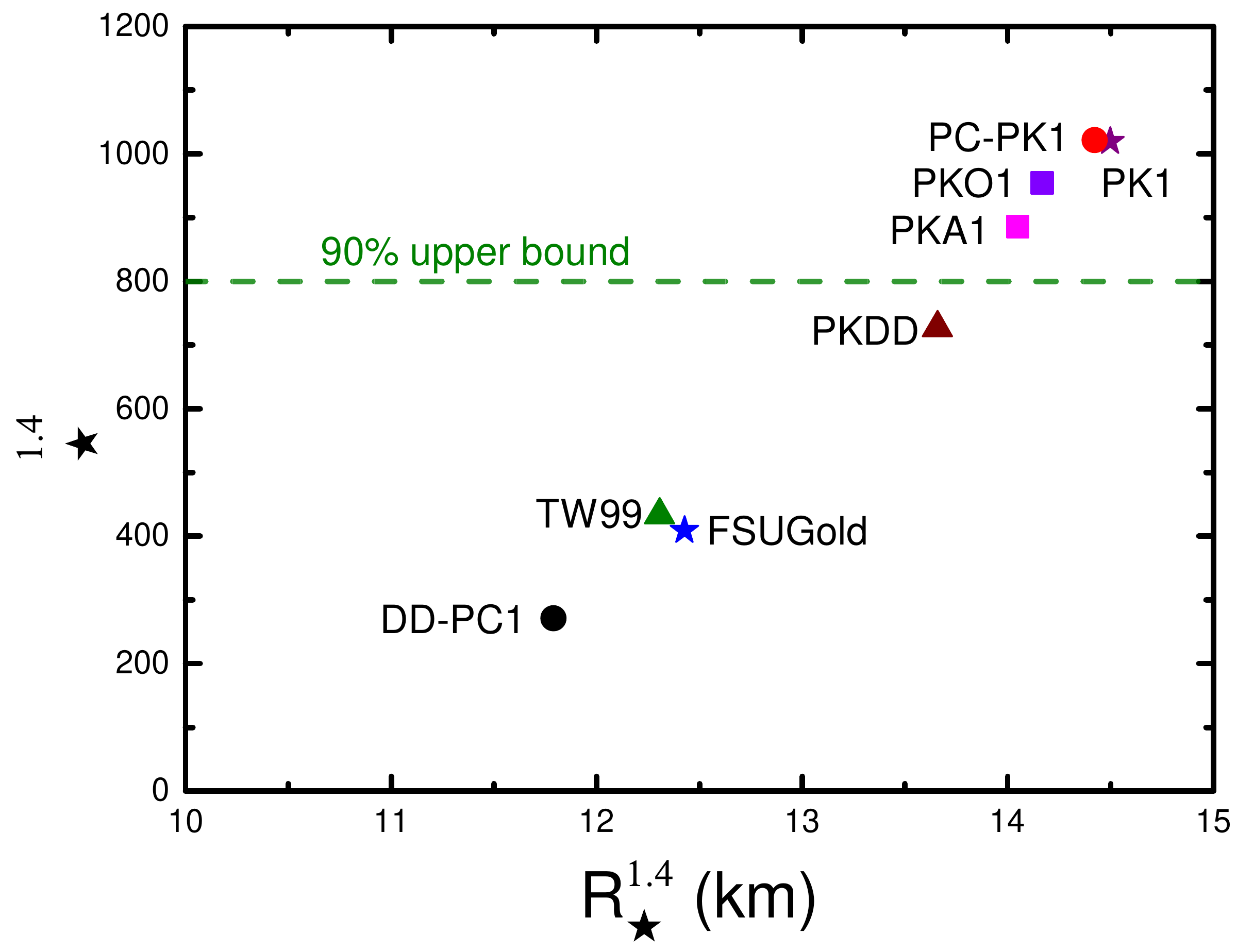}}
  \caption{The dimensionless tidal deformability $\Lambda^{1.4}_{\bigstar}$ of a 1.4~M$_\odot$ neutron star as a function of the corresponding stellar radius. The limit $\Lambda_{\bigstar}^{1.4}\leqslant800$ is deduced from GW170817 data \cite{Abbott2017PRL119.161101}.}\label{Fig:L-R}
\end{figure}

Recently, the analysis of GW170817 data by LIGO-Virgo Collaboration set an upper bound on the dimensionless tidal deformability of a 1.4~M$_\odot$ neutron star, i.e., $\Lambda_{\bigstar}^{1.4}\leqslant800$ with 90\% confidence \cite{Abbott2017PRL119.161101}. After that, a plot of $\Lambda^{1.4}_{\bigstar}$ with respect to the corresponding stellar radius or compactness parameter is investigated explosively by many authors using a more diverse set of models for the nuclear EoS, such as in Ref. \cite{Fattoyev2018PRL120.172702, PhysRevLett.121.091102, PhysRevC.98.035804}, and a strong correlation between these quantities is claimed, accordingly which can be used to constrain the EoS and the symmetry energy at high densities. In Fig. \ref{Fig:L-R}, the similar pattern is also recognized within the selected CDF functionals including PC-RMF ones. Those with relatively strong density dependence of the symmetry energy (shown in Fig.\ref{Fig:symmetry_energy}), namely, PKA1, PKO1, PK1 and PC-PK1, are ruled out by the constraint $\Lambda_{\bigstar}^{1.4}\leqslant800$. For PC-RMF models, there exists an enormous disparity between DD-PC1 and PC-PK1 in predicting the value of $\Lambda^{1.4}_{\bigstar}$. The stringent constraint on $\Lambda_{\bigstar}$ from observational data of gravitational wave event or from microscopic modeling of dense matter EoS \cite{PhysRevLett.121.062701} then makes sense to reduce uncertainty of the dense matter EoS and improving the parameterizing of the PC-RMF models.

\section{Summary}
In this work, the applicability of the relativistic point-coupling models to establish the EoS of neutron star matter and to investigate various neutron star properties has been discussed. The analysis has been done by two selected PC-RMF functionals DD-PC1 and PC-PK1, which correspond to density-dependent and nonlinear types, respectively. In PC-PK1, the predicted pressure of neutron star matter drops down at high densities due to the negative contribution from $\gamma_V$ relevant term, leading to the difficulty of neutron stars approaching to the maximum mass limit. In addition, the divergences between the selected functionals in describing neutron star quantities, such as proton fractions, mass-radius relations, the fraction of crustal moment of inertia and dimensionless tidal deformabilities, are still remarkable, owing to the less constrained behavior of these functionals at high densities. To improve further the description of neutron star physics within the relativistic point-coupling models, it is then suggested that the constraints on the dense matter EoS from modern astronomical observations, such as the tidal-deformabilities taken from gravitational-wave events, should be taken into account during the parameterizing procedure of models.

\section{ACKNOWLEDGMENTS}
We thank James Lattimer, Jeremy Holt, Nobutoshi Yasutake, Jinniu Hu and Lijing Shao for helpful discussions during the Xiamen-CUSTIPENWorkshop on the EOS of Dense Neutron-Rich Matter in the Era of Gravitational Wave Astronomy. This work is partly supported by the National Natural Science Foundation of China (Grant Nos. 11675065 and 11875152).

\nocite{*}
\bibliographystyle{aipnum-cp}%
\bibliography{PC-RMF}%

\begin{thebibliography}{88}%
\makeatletter
\providecommand \@ifxundefined [1]{%
 \@ifx{#1\undefined}
}%
\providecommand \@ifnum [1]{%
 \ifnum #1\expandafter \@firstoftwo
 \else \expandafter \@secondoftwo
 \fi
}%
\providecommand \@ifx [1]{%
 \ifx #1\expandafter \@firstoftwo
 \else \expandafter \@secondoftwo
 \fi
}%
\providecommand \natexlab [1]{#1}%
\providecommand \enquote  [1]{``#1''}%
\providecommand \bibnamefont  [1]{#1}%
\providecommand \bibfnamefont [1]{#1}%
\providecommand \citenamefont [1]{#1}%
\providecommand \href@noop [0]{\@secondoftwo}%
\providecommand \href [0]{\begingroup \@sanitize@url \@href}%
\providecommand \@href[1]{\@@startlink{#1}\@@href}%
\providecommand \@@href[1]{\endgroup#1\@@endlink}%
\providecommand \@sanitize@url [0]{\catcode `\$12\catcode `\&12\catcode
  `\#12\catcode `\^12\catcode `\_12\catcode `\%12\relax}%
\providecommand \@@startlink[1]{}%
\providecommand \@@endlink[0]{}%
\providecommand \url  [0]{\begingroup\@sanitize@url \@url }%
\providecommand \@url [1]{\endgroup\@href {#1}{\urlprefix }}%
\providecommand \urlprefix  [0]{URL }%
\providecommand \Eprint [0]{\href }%
\providecommand \doibase [0]{http://dx.doi.org/}%
\providecommand \selectlanguage [0]{\@gobble}%
\providecommand \bibinfo  [0]{\@secondoftwo}%
\providecommand \bibfield  [0]{\@secondoftwo}%
\providecommand \translation [1]{[#1]}%
\providecommand \BibitemOpen [0]{}%
\providecommand \bibitemStop [0]{}%
\providecommand \bibitemNoStop [0]{.\EOS\space}%
\providecommand \EOS [0]{\spacefactor3000\relax}%
\providecommand \BibitemShut  [1]{\csname bibitem#1\endcsname}%
\let\auto@bib@innerbib\@empty
\bibitem [{\citenamefont {Abbott~et
  al.}(2017{\natexlab{a}})}]{Abbott2017PRL119.161101}%
  \BibitemOpen
  \bibfield  {author} {\bibinfo {author} {\bibfnamefont {B.~P.}\ \bibnamefont
  {Abbott~et al.}} (\bibinfo {collaboration} {LIGO Scientific Collaboration and
  Virgo Collaboration}),\ }\href@noop {} {\bibfield  {journal} {\bibinfo
  {journal} {Phys. Rev. Lett.}\ }\textbf {\bibinfo {volume} {119}},\ p.\
  \bibinfo {pages} {161101} (\bibinfo {year} {2017}{\natexlab{a}})}\BibitemShut
  {NoStop}%
\bibitem [{\citenamefont {Abbott~et al.}(2017{\natexlab{b}})}]{Abbott2017ApJL}%
  \BibitemOpen
  \bibfield  {author} {\bibinfo {author} {\bibfnamefont {B.~P.}\ \bibnamefont
  {Abbott~et al.}},\ }\href@noop {} {\bibfield  {journal} {\bibinfo  {journal}
  {Astrophys. J.}\ }\textbf {\bibinfo {volume} {848}},\ p.\ \bibinfo {pages}
  {L12} (\bibinfo {year} {2017}{\natexlab{b}})}\BibitemShut {NoStop}%
\bibitem [{\citenamefont {Margalit}\ and\ \citenamefont
  {Metzger}(2017)}]{Margalit2017}%
  \BibitemOpen
  \bibfield  {author} {\bibinfo {author} {\bibfnamefont {B.}~\bibnamefont
  {Margalit}}\ and\ \bibinfo {author} {\bibfnamefont {B.~D.}\ \bibnamefont
  {Metzger}},\ }\href@noop {} {\bibfield  {journal} {\bibinfo  {journal}
  {Astrophys. J.}\ }\textbf {\bibinfo {volume} {850}},\ p.\ \bibinfo {pages}
  {L19} (\bibinfo {year} {2017})}\BibitemShut {NoStop}%
\bibitem [{\citenamefont {Rezzolla}, \citenamefont {Most},\ and\ \citenamefont
  {Weih}(2018)}]{Rezzolla2018}%
  \BibitemOpen
  \bibfield  {author} {\bibinfo {author} {\bibfnamefont {L.}~\bibnamefont
  {Rezzolla}}, \bibinfo {author} {\bibfnamefont {E.~R.}\ \bibnamefont {Most}},
  \ and\ \bibinfo {author} {\bibfnamefont {L.~R.}\ \bibnamefont {Weih}},\
  }\href {\doibase 10.3847/2041-8213/aaa401} {\bibfield  {journal} {\bibinfo
  {journal} {Astrophys. J.}\ }\textbf {\bibinfo {volume} {852}},\ p.\ \bibinfo
  {pages} {L25} (\bibinfo {year} {2018})}\BibitemShut {NoStop}%
\bibitem [{\citenamefont {Bauswein}\ \emph {et~al.}(2017)\citenamefont
  {Bauswein}, \citenamefont {Just}, \citenamefont {Janka},\ and\ \citenamefont
  {Stergioulas}}]{Bauswein2017}%
  \BibitemOpen
  \bibfield  {author} {\bibinfo {author} {\bibfnamefont {A.}~\bibnamefont
  {Bauswein}}, \bibinfo {author} {\bibfnamefont {O.}~\bibnamefont {Just}},
  \bibinfo {author} {\bibfnamefont {H.~T.}\ \bibnamefont {Janka}}, \ and\
  \bibinfo {author} {\bibfnamefont {N.}~\bibnamefont {Stergioulas}},\ }\href
  {\doibase 10.3847/2041-8213/aa9994} {\bibfield  {journal} {\bibinfo
  {journal} {Astrophys. J.}\ }\textbf {\bibinfo {volume} {850}},\ p.\ \bibinfo
  {pages} {L34} (\bibinfo {year} {2017})}\BibitemShut {NoStop}%
\bibitem [{\citenamefont {Annala}\ \emph {et~al.}(2018)\citenamefont {Annala},
  \citenamefont {Gorda}, \citenamefont {Kurkela},\ and\ \citenamefont
  {Vuorinen}}]{PhysRevLett.120.172703}%
  \BibitemOpen
  \bibfield  {author} {\bibinfo {author} {\bibfnamefont {E.}~\bibnamefont
  {Annala}}, \bibinfo {author} {\bibfnamefont {T.}~\bibnamefont {Gorda}},
  \bibinfo {author} {\bibfnamefont {A.}~\bibnamefont {Kurkela}}, \ and\
  \bibinfo {author} {\bibfnamefont {A.}~\bibnamefont {Vuorinen}},\ }\href
  {\doibase 10.1103/PhysRevLett.120.172703} {\bibfield  {journal} {\bibinfo
  {journal} {Phys. Rev. Lett.}\ }\textbf {\bibinfo {volume} {120}},\ p.\
  \bibinfo {pages} {172703} (\bibinfo {year} {2018})}\BibitemShut {NoStop}%
\bibitem [{\citenamefont {Abbott~et al.}(2018)}]{PhysRevLett.121.161101}%
  \BibitemOpen
  \bibfield  {author} {\bibinfo {author} {\bibfnamefont {B.~P.}\ \bibnamefont
  {Abbott~et al.}} (\bibinfo {collaboration} {LIGO Scientific Collaboration and
  Virgo Collaboration}),\ }\href {\doibase 10.1103/PhysRevLett.121.161101}
  {\bibfield  {journal} {\bibinfo  {journal} {Phys. Rev. Lett.}\ }\textbf
  {\bibinfo {volume} {121}},\ p.\ \bibinfo {pages} {161101} (\bibinfo {year}
  {2018})}\BibitemShut {NoStop}%
\bibitem [{\citenamefont {Malik}\ \emph {et~al.}(2018)\citenamefont {Malik},
  \citenamefont {Alam}, \citenamefont {Fortin}, \citenamefont {Provid\^encia},
  \citenamefont {Agrawal}, \citenamefont {Jha}, \citenamefont {Kumar},\ and\
  \citenamefont {Patra}}]{PhysRevC.98.035804}%
  \BibitemOpen
  \bibfield  {author} {\bibinfo {author} {\bibfnamefont {T.}~\bibnamefont
  {Malik}}, \bibinfo {author} {\bibfnamefont {N.}~\bibnamefont {Alam}},
  \bibinfo {author} {\bibfnamefont {M.}~\bibnamefont {Fortin}}, \bibinfo
  {author} {\bibfnamefont {C.}~\bibnamefont {Provid\^encia}}, \bibinfo {author}
  {\bibfnamefont {B.~K.}\ \bibnamefont {Agrawal}}, \bibinfo {author}
  {\bibfnamefont {T.~K.}\ \bibnamefont {Jha}}, \bibinfo {author} {\bibfnamefont
  {B.}~\bibnamefont {Kumar}}, \ and\ \bibinfo {author} {\bibfnamefont {S.~K.}\
  \bibnamefont {Patra}},\ }\href {\doibase 10.1103/PhysRevC.98.035804}
  {\bibfield  {journal} {\bibinfo  {journal} {Phys. Rev. C}\ }\textbf {\bibinfo
  {volume} {98}},\ p.\ \bibinfo {pages} {035804} (\bibinfo {year}
  {2018})}\BibitemShut {NoStop}%
\bibitem [{\citenamefont {Zhang}, \citenamefont {Li},\ and\ \citenamefont
  {Xu}(2018)}]{Zhang2018}%
  \BibitemOpen
  \bibfield  {author} {\bibinfo {author} {\bibfnamefont {N.~B.}\ \bibnamefont
  {Zhang}}, \bibinfo {author} {\bibfnamefont {B.~A.}\ \bibnamefont {Li}}, \
  and\ \bibinfo {author} {\bibfnamefont {J.}~\bibnamefont {Xu}},\ }\href
  {\doibase 10.3847/1538-4357/aac027} {\bibfield  {journal} {\bibinfo
  {journal} {Astrophys. J.}\ }\textbf {\bibinfo {volume} {859}},\ p.~\bibinfo
  {pages} {90} (\bibinfo {year} {2018})}\BibitemShut {NoStop}%
\bibitem [{\citenamefont {Bender}, \citenamefont {Heenen},\ and\ \citenamefont
  {Reinhard}(2003)}]{BenderRMP}%
  \BibitemOpen
  \bibfield  {author} {\bibinfo {author} {\bibfnamefont {M.}~\bibnamefont
  {Bender}}, \bibinfo {author} {\bibfnamefont {P.~H.}\ \bibnamefont {Heenen}},
  \ and\ \bibinfo {author} {\bibfnamefont {P.~G.}\ \bibnamefont {Reinhard}},\
  }\href {\doibase 10.1103/RevModPhys.75.121} {\bibfield  {journal} {\bibinfo
  {journal} {Rev. Mod. Phys.}\ }\textbf {\bibinfo {volume} {75}},\ p.\ \bibinfo
  {pages} {121} (\bibinfo {year} {2003})}\BibitemShut {NoStop}%
\bibitem [{\citenamefont {Fayans}\ \emph {et~al.}(2000)\citenamefont {Fayans},
  \citenamefont {Tolokonnikov}, \citenamefont {Trykov},\ and\ \citenamefont
  {Zawischa}}]{FAYANS200049}%
  \BibitemOpen
  \bibfield  {author} {\bibinfo {author} {\bibfnamefont {S.}~\bibnamefont
  {Fayans}}, \bibinfo {author} {\bibfnamefont {S.}~\bibnamefont
  {Tolokonnikov}}, \bibinfo {author} {\bibfnamefont {E.}~\bibnamefont
  {Trykov}}, \ and\ \bibinfo {author} {\bibfnamefont {D.}~\bibnamefont
  {Zawischa}},\ }\href {\doibase https://doi.org/10.1016/S0375-9474(00)00192-5}
  {\bibfield  {journal} {\bibinfo  {journal} {Nucl. Phys. A}\ }\textbf
  {\bibinfo {volume} {676}},\ p.~\bibinfo {pages} {49} (\bibinfo {year}
  {2000})}\BibitemShut {NoStop}%
\bibitem [{\citenamefont {Vretenar}\ \emph {et~al.}(2005)\citenamefont
  {Vretenar}, \citenamefont {Afanasjev}, \citenamefont {Lalazissis},\ and\
  \citenamefont {Ring}}]{VRETENAR2005101}%
  \BibitemOpen
  \bibfield  {author} {\bibinfo {author} {\bibfnamefont {D.}~\bibnamefont
  {Vretenar}}, \bibinfo {author} {\bibfnamefont {A.}~\bibnamefont {Afanasjev}},
  \bibinfo {author} {\bibfnamefont {G.}~\bibnamefont {Lalazissis}}, \ and\
  \bibinfo {author} {\bibfnamefont {P.}~\bibnamefont {Ring}},\ }\href {\doibase
  https://doi.org/10.1016/j.physrep.2004.10.001} {\bibfield  {journal}
  {\bibinfo  {journal} {Phys. Rep.}\ }\textbf {\bibinfo {volume} {409}},\ p.\
  \bibinfo {pages} {101} (\bibinfo {year} {2005})}\BibitemShut {NoStop}%
\bibitem [{\citenamefont {Meng}\ \emph {et~al.}(2006)\citenamefont {Meng},
  \citenamefont {Toki}, \citenamefont {Zhou}, \citenamefont {Zhang},
  \citenamefont {Long},\ and\ \citenamefont {Geng}}]{MENG2006470}%
  \BibitemOpen
  \bibfield  {author} {\bibinfo {author} {\bibfnamefont {J.}~\bibnamefont
  {Meng}}, \bibinfo {author} {\bibfnamefont {H.}~\bibnamefont {Toki}}, \bibinfo
  {author} {\bibfnamefont {S.~G.}\ \bibnamefont {Zhou}}, \bibinfo {author}
  {\bibfnamefont {S.~Q.}\ \bibnamefont {Zhang}}, \bibinfo {author}
  {\bibfnamefont {W.~H.}\ \bibnamefont {Long}}, \ and\ \bibinfo {author}
  {\bibfnamefont {L.~S.}\ \bibnamefont {Geng}},\ }\href {\doibase
  https://doi.org/10.1016/j.ppnp.2005.06.001} {\bibfield  {journal} {\bibinfo
  {journal} {Prog. Part. Nucl. Phys.}\ }\textbf {\bibinfo {volume} {57}},\ p.\
  \bibinfo {pages} {470} (\bibinfo {year} {2006})}\BibitemShut {NoStop}%
\bibitem [{\citenamefont {Ginocchio}(2005)}]{GINOCCHIO2005165}%
  \BibitemOpen
  \bibfield  {author} {\bibinfo {author} {\bibfnamefont {J.~N.}\ \bibnamefont
  {Ginocchio}},\ }\href {\doibase
  https://doi.org/10.1016/j.physrep.2005.04.003} {\bibfield  {journal}
  {\bibinfo  {journal} {Phys. Rep.}\ }\textbf {\bibinfo {volume} {414}},\ p.\
  \bibinfo {pages} {165} (\bibinfo {year} {2005})}\BibitemShut {NoStop}%
\bibitem [{\citenamefont {Liang}, \citenamefont {Meng},\ and\ \citenamefont
  {Zhou}(2015)}]{LIANG20151}%
  \BibitemOpen
  \bibfield  {author} {\bibinfo {author} {\bibfnamefont {H.~Z.}\ \bibnamefont
  {Liang}}, \bibinfo {author} {\bibfnamefont {J.}~\bibnamefont {Meng}}, \ and\
  \bibinfo {author} {\bibfnamefont {S.~G.}\ \bibnamefont {Zhou}},\ }\href
  {\doibase https://doi.org/10.1016/j.physrep.2014.12.005} {\bibfield
  {journal} {\bibinfo  {journal} {Phys. Rep.}\ }\textbf {\bibinfo {volume}
  {570}},\ p.~\bibinfo {pages} {1} (\bibinfo {year} {2015})}\BibitemShut
  {NoStop}%
\bibitem [{\citenamefont {Boguta}\ and\ \citenamefont
  {Stocker}(1983)}]{BOGUTA1983289}%
  \BibitemOpen
  \bibfield  {author} {\bibinfo {author} {\bibfnamefont {J.}~\bibnamefont
  {Boguta}}\ and\ \bibinfo {author} {\bibfnamefont {H.}~\bibnamefont
  {Stocker}},\ }\href {\doibase https://doi.org/10.1016/0370-2693(83)90446-X}
  {\bibfield  {journal} {\bibinfo  {journal} {Phys. Lett. B}\ }\textbf
  {\bibinfo {volume} {120}},\ p.\ \bibinfo {pages} {289} (\bibinfo {year}
  {1983})}\BibitemShut {NoStop}%
\bibitem [{\citenamefont {Typel}\ and\ \citenamefont
  {Wolter}(1999{\natexlab{a}})}]{TYPEL1999331}%
  \BibitemOpen
  \bibfield  {author} {\bibinfo {author} {\bibfnamefont {S.}~\bibnamefont
  {Typel}}\ and\ \bibinfo {author} {\bibfnamefont {H.~H.}\ \bibnamefont
  {Wolter}},\ }\href {\doibase https://doi.org/10.1016/S0375-9474(99)00310-3}
  {\bibfield  {journal} {\bibinfo  {journal} {Nucl. Phys. A}\ }\textbf
  {\bibinfo {volume} {656}},\ p.\ \bibinfo {pages} {331} (\bibinfo {year}
  {1999}{\natexlab{a}})}\BibitemShut {NoStop}%
\bibitem [{\citenamefont {Long}, \citenamefont {Van~Giai},\ and\ \citenamefont
  {Meng}(2006)}]{Long2006PLB640.150}%
  \BibitemOpen
  \bibfield  {author} {\bibinfo {author} {\bibfnamefont {W.~H.}\ \bibnamefont
  {Long}}, \bibinfo {author} {\bibfnamefont {N.}~\bibnamefont {Van~Giai}}, \
  and\ \bibinfo {author} {\bibfnamefont {J.}~\bibnamefont {Meng}},\ }\href@noop
  {} {\bibfield  {journal} {\bibinfo  {journal} {Phys. Lett. B}\ }\textbf
  {\bibinfo {volume} {640}},\ p.\ \bibinfo {pages} {150} (\bibinfo {year}
  {2006})}\BibitemShut {NoStop}%
\bibitem [{\citenamefont {Long}\ \emph {et~al.}(2007)\citenamefont {Long},
  \citenamefont {Sagawa}, \citenamefont {Van~Giai},\ and\ \citenamefont
  {Meng}}]{Long2007PRC76.034314}%
  \BibitemOpen
  \bibfield  {author} {\bibinfo {author} {\bibfnamefont {W.~H.}\ \bibnamefont
  {Long}}, \bibinfo {author} {\bibfnamefont {H.}~\bibnamefont {Sagawa}},
  \bibinfo {author} {\bibfnamefont {N.}~\bibnamefont {Van~Giai}}, \ and\
  \bibinfo {author} {\bibfnamefont {J.}~\bibnamefont {Meng}},\ }\href@noop {}
  {\bibfield  {journal} {\bibinfo  {journal} {Phys. Rev. C}\ }\textbf {\bibinfo
  {volume} {76}},\ p.\ \bibinfo {pages} {034314} (\bibinfo {year}
  {2007})}\BibitemShut {NoStop}%
\bibitem [{\citenamefont {Jiang}\ \emph
  {et~al.}(2015{\natexlab{a}})\citenamefont {Jiang}, \citenamefont {Yang},
  \citenamefont {Sun}, \citenamefont {Long},\ and\ \citenamefont
  {Gu}}]{Jiang2015PRC91.034326}%
  \BibitemOpen
  \bibfield  {author} {\bibinfo {author} {\bibfnamefont {L.~J.}\ \bibnamefont
  {Jiang}}, \bibinfo {author} {\bibfnamefont {S.}~\bibnamefont {Yang}},
  \bibinfo {author} {\bibfnamefont {B.~Y.}\ \bibnamefont {Sun}}, \bibinfo
  {author} {\bibfnamefont {W.~H.}\ \bibnamefont {Long}}, \ and\ \bibinfo
  {author} {\bibfnamefont {H.~Q.}\ \bibnamefont {Gu}},\ }\href@noop {}
  {\bibfield  {journal} {\bibinfo  {journal} {Phys. Rev. C}\ }\textbf {\bibinfo
  {volume} {91}},\ p.\ \bibinfo {pages} {034326} (\bibinfo {year}
  {2015}{\natexlab{a}})}\BibitemShut {NoStop}%
\bibitem [{\citenamefont {Zong}\ and\ \citenamefont
  {Sun}(2018)}]{Zong2018CPC42.24101}%
  \BibitemOpen
  \bibfield  {author} {\bibinfo {author} {\bibfnamefont {Y.~Y.}\ \bibnamefont
  {Zong}}\ and\ \bibinfo {author} {\bibfnamefont {B.~Y.}\ \bibnamefont {Sun}},\
  }\href@noop {} {\bibfield  {journal} {\bibinfo  {journal} {Chin. Phys. C}\
  }\textbf {\bibinfo {volume} {42}},\ p.\ \bibinfo {pages} {24101} (\bibinfo
  {year} {2018})}\BibitemShut {NoStop}%
\bibitem [{\citenamefont {Wang}\ \emph {et~al.}(2018)\citenamefont {Wang},
  \citenamefont {Zhao}, \citenamefont {Liang},\ and\ \citenamefont
  {Long}}]{Wang2018PRC98.034313}%
  \BibitemOpen
  \bibfield  {author} {\bibinfo {author} {\bibfnamefont {Z.~H.}\ \bibnamefont
  {Wang}}, \bibinfo {author} {\bibfnamefont {Q.}~\bibnamefont {Zhao}}, \bibinfo
  {author} {\bibfnamefont {H.~Z.}\ \bibnamefont {Liang}}, \ and\ \bibinfo
  {author} {\bibfnamefont {W.~H.}\ \bibnamefont {Long}},\ }\href@noop {}
  {\bibfield  {journal} {\bibinfo  {journal} {Phys. Rev. C}\ }\textbf {\bibinfo
  {volume} {98}},\ p.\ \bibinfo {pages} {034313} (\bibinfo {year}
  {2018})}\BibitemShut {NoStop}%
\bibitem [{\citenamefont {Vretenar}, \citenamefont {Nik\ifmmode \check{s}\else
  \v{s}\fi{}i\ifmmode~\acute{c}\else \'{c}\fi{}},\ and\ \citenamefont
  {Ring}(2003)}]{Vretenar2003PRC68.024310}%
  \BibitemOpen
  \bibfield  {author} {\bibinfo {author} {\bibfnamefont {D.}~\bibnamefont
  {Vretenar}}, \bibinfo {author} {\bibfnamefont {T.}~\bibnamefont {Nik\ifmmode
  \check{s}\else \v{s}\fi{}i\ifmmode~\acute{c}\else \'{c}\fi{}}}, \ and\
  \bibinfo {author} {\bibfnamefont {P.}~\bibnamefont {Ring}},\ }\href@noop {}
  {\bibfield  {journal} {\bibinfo  {journal} {Phys. Rev. C}\ }\textbf {\bibinfo
  {volume} {68}},\ p.\ \bibinfo {pages} {024310} (\bibinfo {year}
  {2003})}\BibitemShut {NoStop}%
\bibitem [{\citenamefont {Ban}\ \emph {et~al.}(2004)\citenamefont {Ban},
  \citenamefont {Li}, \citenamefont {Zhang}, \citenamefont {Jia}, \citenamefont
  {Sang},\ and\ \citenamefont {Meng}}]{Ban2004PRC69.045805}%
  \BibitemOpen
  \bibfield  {author} {\bibinfo {author} {\bibfnamefont {S.~F.}\ \bibnamefont
  {Ban}}, \bibinfo {author} {\bibfnamefont {J.}~\bibnamefont {Li}}, \bibinfo
  {author} {\bibfnamefont {S.~Q.}\ \bibnamefont {Zhang}}, \bibinfo {author}
  {\bibfnamefont {H.~Y.}\ \bibnamefont {Jia}}, \bibinfo {author} {\bibfnamefont
  {J.~P.}\ \bibnamefont {Sang}}, \ and\ \bibinfo {author} {\bibfnamefont
  {J.}~\bibnamefont {Meng}},\ }\href {\doibase 10.1103/PhysRevC.69.045805}
  {\bibfield  {journal} {\bibinfo  {journal} {Phys. Rev. C}\ }\textbf {\bibinfo
  {volume} {69}},\ p.\ \bibinfo {pages} {045805} (\bibinfo {year}
  {2004})}\BibitemShut {NoStop}%
\bibitem [{\citenamefont {Chen}, \citenamefont {Ko},\ and\ \citenamefont
  {Li}(2007)}]{Chen2007PRC76.054316}%
  \BibitemOpen
  \bibfield  {author} {\bibinfo {author} {\bibfnamefont {L.~W.}\ \bibnamefont
  {Chen}}, \bibinfo {author} {\bibfnamefont {C.~M.}\ \bibnamefont {Ko}}, \ and\
  \bibinfo {author} {\bibfnamefont {B.~A.}\ \bibnamefont {Li}},\ }\href@noop {}
  {\bibfield  {journal} {\bibinfo  {journal} {Phys. Rev. C}\ }\textbf {\bibinfo
  {volume} {76}},\ p.\ \bibinfo {pages} {054316} (\bibinfo {year}
  {2007})}\BibitemShut {NoStop}%
\bibitem [{\citenamefont {Sun}\ \emph {et~al.}(2008)\citenamefont {Sun},
  \citenamefont {Long}, \citenamefont {Meng},\ and\ \citenamefont
  {Lombardo}}]{Sun2008PRC78.065805}%
  \BibitemOpen
  \bibfield  {author} {\bibinfo {author} {\bibfnamefont {B.~Y.}\ \bibnamefont
  {Sun}}, \bibinfo {author} {\bibfnamefont {W.~H.}\ \bibnamefont {Long}},
  \bibinfo {author} {\bibfnamefont {J.}~\bibnamefont {Meng}}, \ and\ \bibinfo
  {author} {\bibfnamefont {U.}~\bibnamefont {Lombardo}},\ }\href@noop {}
  {\bibfield  {journal} {\bibinfo  {journal} {Phys. Rev. C}\ }\textbf {\bibinfo
  {volume} {78}},\ p.\ \bibinfo {pages} {065805} (\bibinfo {year}
  {2008})}\BibitemShut {NoStop}%
\bibitem [{\citenamefont {Sun}, \citenamefont {Toki},\ and\ \citenamefont
  {Meng}(2010)}]{BySun2010}%
  \BibitemOpen
  \bibfield  {author} {\bibinfo {author} {\bibfnamefont {B.~Y.}\ \bibnamefont
  {Sun}}, \bibinfo {author} {\bibfnamefont {H.}~\bibnamefont {Toki}}, \ and\
  \bibinfo {author} {\bibfnamefont {J.}~\bibnamefont {Meng}},\ }\href {\doibase
  DOI: 10.1016/j.physletb.2009.11.065} {\bibfield  {journal} {\bibinfo
  {journal} {Phys. Lett. B}\ }\textbf {\bibinfo {volume} {683}},\ p.\ \bibinfo
  {pages} {134} (\bibinfo {year} {2010})}\BibitemShut {NoStop}%
\bibitem [{\citenamefont {Long}\ \emph {et~al.}(2012)\citenamefont {Long},
  \citenamefont {Sun}, \citenamefont {Hagino},\ and\ \citenamefont
  {Sagawa}}]{Long2012PRC85.025806}%
  \BibitemOpen
  \bibfield  {author} {\bibinfo {author} {\bibfnamefont {W.~H.}\ \bibnamefont
  {Long}}, \bibinfo {author} {\bibfnamefont {B.~Y.}\ \bibnamefont {Sun}},
  \bibinfo {author} {\bibfnamefont {K.}~\bibnamefont {Hagino}}, \ and\ \bibinfo
  {author} {\bibfnamefont {H.}~\bibnamefont {Sagawa}},\ }\href@noop {}
  {\bibfield  {journal} {\bibinfo  {journal} {Phys. Rev. C}\ }\textbf {\bibinfo
  {volume} {85}},\ p.\ \bibinfo {pages} {025806} (\bibinfo {year}
  {2012})}\BibitemShut {NoStop}%
\bibitem [{\citenamefont {Jiang}\ \emph
  {et~al.}(2015{\natexlab{b}})\citenamefont {Jiang}, \citenamefont {Yang},
  \citenamefont {Dong},\ and\ \citenamefont {Long}}]{Jiang2015PRC91.025802}%
  \BibitemOpen
  \bibfield  {author} {\bibinfo {author} {\bibfnamefont {L.~J.}\ \bibnamefont
  {Jiang}}, \bibinfo {author} {\bibfnamefont {S.}~\bibnamefont {Yang}},
  \bibinfo {author} {\bibfnamefont {J.~M.}\ \bibnamefont {Dong}}, \ and\
  \bibinfo {author} {\bibfnamefont {W.~H.}\ \bibnamefont {Long}},\ }\href@noop
  {} {\bibfield  {journal} {\bibinfo  {journal} {Phys. Rev. C}\ }\textbf
  {\bibinfo {volume} {91}},\ p.\ \bibinfo {pages} {025802} (\bibinfo {year}
  {2015}{\natexlab{b}})}\BibitemShut {NoStop}%
\bibitem [{\citenamefont {Zhao}, \citenamefont {Sun},\ and\ \citenamefont
  {Long}(2015)}]{Zhao2015JPG42.095101}%
  \BibitemOpen
  \bibfield  {author} {\bibinfo {author} {\bibfnamefont {Q.}~\bibnamefont
  {Zhao}}, \bibinfo {author} {\bibfnamefont {B.~Y.}\ \bibnamefont {Sun}}, \
  and\ \bibinfo {author} {\bibfnamefont {W.~H.}\ \bibnamefont {Long}},\
  }\href@noop {} {\bibfield  {journal} {\bibinfo  {journal} {J. Phys. G: Nucl.
  Part. Phys.}\ }\textbf {\bibinfo {volume} {42}},\ p.\ \bibinfo {pages}
  {095101} (\bibinfo {year} {2015})}\BibitemShut {NoStop}%
\bibitem [{\citenamefont {Liu}\ \emph {et~al.}(2018)\citenamefont {Liu},
  \citenamefont {Qian}, \citenamefont {Xing}, \citenamefont {Niu},\ and\
  \citenamefont {Sun}}]{Liu2018PRC97.025801}%
  \BibitemOpen
  \bibfield  {author} {\bibinfo {author} {\bibfnamefont {Z.~W.}\ \bibnamefont
  {Liu}}, \bibinfo {author} {\bibfnamefont {Z.}~\bibnamefont {Qian}}, \bibinfo
  {author} {\bibfnamefont {R.~Y.}\ \bibnamefont {Xing}}, \bibinfo {author}
  {\bibfnamefont {J.~R.}\ \bibnamefont {Niu}}, \ and\ \bibinfo {author}
  {\bibfnamefont {B.~Y.}\ \bibnamefont {Sun}},\ }\href@noop {} {\bibfield
  {journal} {\bibinfo  {journal} {Phys. Rev. C}\ }\textbf {\bibinfo {volume}
  {97}},\ p.\ \bibinfo {pages} {025801} (\bibinfo {year} {2018})}\BibitemShut
  {NoStop}%
\bibitem [{\citenamefont {Nikolaus}, \citenamefont {Hoch},\ and\ \citenamefont
  {Madland}(1992)}]{Nikolaus1992PRC46.1757}%
  \BibitemOpen
  \bibfield  {author} {\bibinfo {author} {\bibfnamefont {B.~A.}\ \bibnamefont
  {Nikolaus}}, \bibinfo {author} {\bibfnamefont {T.}~\bibnamefont {Hoch}}, \
  and\ \bibinfo {author} {\bibfnamefont {D.~G.}\ \bibnamefont {Madland}},\
  }\href@noop {} {\bibfield  {journal} {\bibinfo  {journal} {Phys. Rev. C}\
  }\textbf {\bibinfo {volume} {46}},\ p.\ \bibinfo {pages} {1757} (\bibinfo
  {year} {1992})}\BibitemShut {NoStop}%
\bibitem [{\citenamefont {B\"urvenich}\ \emph {et~al.}(2002)\citenamefont
  {B\"urvenich}, \citenamefont {Madland}, \citenamefont {Maruhn},\ and\
  \citenamefont {Reinhard}}]{Burvenich2002PRC65.044308}%
  \BibitemOpen
  \bibfield  {author} {\bibinfo {author} {\bibfnamefont {T.}~\bibnamefont
  {B\"urvenich}}, \bibinfo {author} {\bibfnamefont {D.~G.}\ \bibnamefont
  {Madland}}, \bibinfo {author} {\bibfnamefont {J.~A.}\ \bibnamefont {Maruhn}},
  \ and\ \bibinfo {author} {\bibfnamefont {P.~G.}\ \bibnamefont {Reinhard}},\
  }\href@noop {} {\bibfield  {journal} {\bibinfo  {journal} {Phys. Rev. C}\
  }\textbf {\bibinfo {volume} {65}},\ p.\ \bibinfo {pages} {044308} (\bibinfo
  {year} {2002})}\BibitemShut {NoStop}%
\bibitem [{\citenamefont {Zhao}\ \emph {et~al.}(2010)\citenamefont {Zhao},
  \citenamefont {Li}, \citenamefont {Yao},\ and\ \citenamefont
  {Meng}}]{Zhao2010PRC82.054319}%
  \BibitemOpen
  \bibfield  {author} {\bibinfo {author} {\bibfnamefont {P.~W.}\ \bibnamefont
  {Zhao}}, \bibinfo {author} {\bibfnamefont {Z.~P.}\ \bibnamefont {Li}},
  \bibinfo {author} {\bibfnamefont {J.~M.}\ \bibnamefont {Yao}}, \ and\
  \bibinfo {author} {\bibfnamefont {J.}~\bibnamefont {Meng}},\ }\href@noop {}
  {\bibfield  {journal} {\bibinfo  {journal} {Phys. Rev. C}\ }\textbf {\bibinfo
  {volume} {82}},\ p.\ \bibinfo {pages} {054319} (\bibinfo {year}
  {2010})}\BibitemShut {NoStop}%
\bibitem [{\citenamefont {Finelli}\ \emph {et~al.}(2006)\citenamefont
  {Finelli}, \citenamefont {Kaiser}, \citenamefont {Vretenar},\ and\
  \citenamefont {Weise}}]{Finelli2006NPA770.1}%
  \BibitemOpen
  \bibfield  {author} {\bibinfo {author} {\bibfnamefont {P.}~\bibnamefont
  {Finelli}}, \bibinfo {author} {\bibfnamefont {N.}~\bibnamefont {Kaiser}},
  \bibinfo {author} {\bibfnamefont {D.}~\bibnamefont {Vretenar}}, \ and\
  \bibinfo {author} {\bibfnamefont {W.}~\bibnamefont {Weise}},\ }\href@noop {}
  {\bibfield  {journal} {\bibinfo  {journal} {Nucl. Phys. A}\ }\textbf
  {\bibinfo {volume} {770}},\ p.~\bibinfo {pages} {1} (\bibinfo {year}
  {2006})}\BibitemShut {NoStop}%
\bibitem [{\citenamefont {Nik\ifmmode \check{s}\else
  \v{s}\fi{}i\ifmmode~\acute{c}\else \'{c}\fi{}}, \citenamefont {Vretenar},\
  and\ \citenamefont {Ring}(2008)}]{Niksic2008PRC78.034318}%
  \BibitemOpen
  \bibfield  {author} {\bibinfo {author} {\bibfnamefont {T.}~\bibnamefont
  {Nik\ifmmode \check{s}\else \v{s}\fi{}i\ifmmode~\acute{c}\else \'{c}\fi{}}},
  \bibinfo {author} {\bibfnamefont {D.}~\bibnamefont {Vretenar}}, \ and\
  \bibinfo {author} {\bibfnamefont {P.}~\bibnamefont {Ring}},\ }\href@noop {}
  {\bibfield  {journal} {\bibinfo  {journal} {Phys. Rev. C}\ }\textbf {\bibinfo
  {volume} {78}},\ p.\ \bibinfo {pages} {034318} (\bibinfo {year}
  {2008})}\BibitemShut {NoStop}%
\bibitem [{\citenamefont {Zhao}\ and\ \citenamefont
  {Li}(2018)}]{Zhao2018IJMPE27.1830007}%
  \BibitemOpen
  \bibfield  {author} {\bibinfo {author} {\bibfnamefont {P.~W.}\ \bibnamefont
  {Zhao}}\ and\ \bibinfo {author} {\bibfnamefont {Z.~P.}\ \bibnamefont {Li}},\
  }\href@noop {} {\bibfield  {journal} {\bibinfo  {journal} {Int. J. Mod. Phys.
  E}\ }\textbf {\bibinfo {volume} {27}},\ p.\ \bibinfo {pages} {1830007}
  (\bibinfo {year} {2018})}\BibitemShut {NoStop}%
\bibitem [{\citenamefont {Sun}, \citenamefont {Zhao},\ and\ \citenamefont
  {Meng}(2011)}]{Sun2011SCPMA54.210}%
  \BibitemOpen
  \bibfield  {author} {\bibinfo {author} {\bibfnamefont {B.~H.}\ \bibnamefont
  {Sun}}, \bibinfo {author} {\bibfnamefont {P.~W.}\ \bibnamefont {Zhao}}, \
  and\ \bibinfo {author} {\bibfnamefont {J.}~\bibnamefont {Meng}},\ }\href@noop
  {} {\bibfield  {journal} {\bibinfo  {journal} {Sci. China Phys. Mech.
  Astron.}\ }\textbf {\bibinfo {volume} {54}},\ p.\ \bibinfo {pages} {210}
  (\bibinfo {year} {2011})}\BibitemShut {NoStop}%
\bibitem [{\citenamefont {Lu}, \citenamefont {Zhao},\ and\ \citenamefont
  {Zhou}(2012)}]{Lu2012PRC85.011301}%
  \BibitemOpen
  \bibfield  {author} {\bibinfo {author} {\bibfnamefont {B.~N.}\ \bibnamefont
  {Lu}}, \bibinfo {author} {\bibfnamefont {E.~G.}\ \bibnamefont {Zhao}}, \ and\
  \bibinfo {author} {\bibfnamefont {S.~G.}\ \bibnamefont {Zhou}},\ }\href@noop
  {} {\bibfield  {journal} {\bibinfo  {journal} {Phys. Rev. C}\ }\textbf
  {\bibinfo {volume} {85}},\ p.\ \bibinfo {pages} {011301} (\bibinfo {year}
  {2012})}\BibitemShut {NoStop}%
\bibitem [{\citenamefont {Zhao}, \citenamefont {Zhang},\ and\ \citenamefont
  {Meng}(2014)}]{Zhao2014PRC89.011301}%
  \BibitemOpen
  \bibfield  {author} {\bibinfo {author} {\bibfnamefont {P.~W.}\ \bibnamefont
  {Zhao}}, \bibinfo {author} {\bibfnamefont {S.~Q.}\ \bibnamefont {Zhang}}, \
  and\ \bibinfo {author} {\bibfnamefont {J.}~\bibnamefont {Meng}},\ }\href@noop
  {} {\bibfield  {journal} {\bibinfo  {journal} {Phys. Rev. C}\ }\textbf
  {\bibinfo {volume} {89}},\ p.\ \bibinfo {pages} {011301} (\bibinfo {year}
  {2014})}\BibitemShut {NoStop}%
\bibitem [{\citenamefont {Li}\ \emph {et~al.}(2012)\citenamefont {Li},
  \citenamefont {Li}, \citenamefont {Xiang}, \citenamefont {Yao},\ and\
  \citenamefont {Meng}}]{Li2012PLB717.470}%
  \BibitemOpen
  \bibfield  {author} {\bibinfo {author} {\bibfnamefont {Z.~P.}\ \bibnamefont
  {Li}}, \bibinfo {author} {\bibfnamefont {C.~Y.}\ \bibnamefont {Li}}, \bibinfo
  {author} {\bibfnamefont {J.}~\bibnamefont {Xiang}}, \bibinfo {author}
  {\bibfnamefont {J.~M.}\ \bibnamefont {Yao}}, \ and\ \bibinfo {author}
  {\bibfnamefont {J.}~\bibnamefont {Meng}},\ }\href@noop {} {\bibfield
  {journal} {\bibinfo  {journal} {Phys. Lett. B}\ }\textbf {\bibinfo {volume}
  {717}},\ p.\ \bibinfo {pages} {470} (\bibinfo {year} {2012})}\BibitemShut
  {NoStop}%
\bibitem [{\citenamefont {Meng}\ \emph {et~al.}(2013)\citenamefont {Meng},
  \citenamefont {Peng}, \citenamefont {Zhang},\ and\ \citenamefont
  {Zhao}}]{Meng2013FP8.55}%
  \BibitemOpen
  \bibfield  {author} {\bibinfo {author} {\bibfnamefont {J.}~\bibnamefont
  {Meng}}, \bibinfo {author} {\bibfnamefont {J.}~\bibnamefont {Peng}}, \bibinfo
  {author} {\bibfnamefont {S.~Q.}\ \bibnamefont {Zhang}}, \ and\ \bibinfo
  {author} {\bibfnamefont {P.~W.}\ \bibnamefont {Zhao}},\ }\href@noop {}
  {\bibfield  {journal} {\bibinfo  {journal} {Front. Phys.}\ }\textbf {\bibinfo
  {volume} {8}},\ p.~\bibinfo {pages} {79} (\bibinfo {year}
  {2013})}\BibitemShut {NoStop}%
\bibitem [{\citenamefont {Zhao}\ \emph
  {et~al.}(2011{\natexlab{a}})\citenamefont {Zhao}, \citenamefont {Zhang},
  \citenamefont {Peng}, \citenamefont {Liang}, \citenamefont {Ring},\ and\
  \citenamefont {Meng}}]{Zhao2011PLB699.181}%
  \BibitemOpen
  \bibfield  {author} {\bibinfo {author} {\bibfnamefont {P.~W.}\ \bibnamefont
  {Zhao}}, \bibinfo {author} {\bibfnamefont {S.~Q.}\ \bibnamefont {Zhang}},
  \bibinfo {author} {\bibfnamefont {J.}~\bibnamefont {Peng}}, \bibinfo {author}
  {\bibfnamefont {H.~Z.}\ \bibnamefont {Liang}}, \bibinfo {author}
  {\bibfnamefont {P.}~\bibnamefont {Ring}}, \ and\ \bibinfo {author}
  {\bibfnamefont {J.}~\bibnamefont {Meng}},\ }\href@noop {} {\bibfield
  {journal} {\bibinfo  {journal} {Phys. Lett. B}\ }\textbf {\bibinfo {volume}
  {699}},\ p.\ \bibinfo {pages} {181} (\bibinfo {year}
  {2011}{\natexlab{a}})}\BibitemShut {NoStop}%
\bibitem [{\citenamefont {Zhao}\ \emph
  {et~al.}(2011{\natexlab{b}})\citenamefont {Zhao}, \citenamefont {Peng},
  \citenamefont {Liang}, \citenamefont {Ring},\ and\ \citenamefont
  {Meng}}]{Zhao2011PRL107.122501}%
  \BibitemOpen
  \bibfield  {author} {\bibinfo {author} {\bibfnamefont {P.~W.}\ \bibnamefont
  {Zhao}}, \bibinfo {author} {\bibfnamefont {J.}~\bibnamefont {Peng}}, \bibinfo
  {author} {\bibfnamefont {H.~Z.}\ \bibnamefont {Liang}}, \bibinfo {author}
  {\bibfnamefont {P.}~\bibnamefont {Ring}}, \ and\ \bibinfo {author}
  {\bibfnamefont {J.}~\bibnamefont {Meng}},\ }\href@noop {} {\bibfield
  {journal} {\bibinfo  {journal} {Phys. Rev. Lett.}\ }\textbf {\bibinfo
  {volume} {107}},\ p.\ \bibinfo {pages} {122501} (\bibinfo {year}
  {2011}{\natexlab{b}})}\BibitemShut {NoStop}%
\bibitem [{\citenamefont {Zhao}(2017)}]{Zhao2017PLB773.1}%
  \BibitemOpen
  \bibfield  {author} {\bibinfo {author} {\bibfnamefont {P.~W.}\ \bibnamefont
  {Zhao}},\ }\href@noop {} {\bibfield  {journal} {\bibinfo  {journal} {Phys.
  Lett. B}\ }\textbf {\bibinfo {volume} {773}},\ p.~\bibinfo {pages} {1}
  (\bibinfo {year} {2017})}\BibitemShut {NoStop}%
\bibitem [{\citenamefont {Miyatsu}, \citenamefont {Katayama},\ and\
  \citenamefont {Saito}(2012)}]{Miyatsu2012PLB709.242}%
  \BibitemOpen
  \bibfield  {author} {\bibinfo {author} {\bibfnamefont {T.}~\bibnamefont
  {Miyatsu}}, \bibinfo {author} {\bibfnamefont {T.}~\bibnamefont {Katayama}}, \
  and\ \bibinfo {author} {\bibfnamefont {K.}~\bibnamefont {Saito}},\
  }\href@noop {} {\bibfield  {journal} {\bibinfo  {journal} {Phys. Lett. B}\
  }\textbf {\bibinfo {volume} {709}},\ p.\ \bibinfo {pages} {242} (\bibinfo
  {year} {2012})}\BibitemShut {NoStop}%
\bibitem [{\citenamefont {Miyatsu}, \citenamefont {Cheoun},\ and\ \citenamefont
  {Saito}(2015)}]{Miyatsu2015AJ813.135}%
  \BibitemOpen
  \bibfield  {author} {\bibinfo {author} {\bibfnamefont {T.}~\bibnamefont
  {Miyatsu}}, \bibinfo {author} {\bibfnamefont {M.~K.}\ \bibnamefont {Cheoun}},
  \ and\ \bibinfo {author} {\bibfnamefont {K.}~\bibnamefont {Saito}},\
  }\href@noop {} {\bibfield  {journal} {\bibinfo  {journal} {Astrophys. J.}\
  }\textbf {\bibinfo {volume} {813}},\ p.\ \bibinfo {pages} {135} (\bibinfo
  {year} {2015})}\BibitemShut {NoStop}%
\bibitem [{\citenamefont {Zhu}\ \emph {et~al.}(2016)\citenamefont {Zhu},
  \citenamefont {Li}, \citenamefont {Hu},\ and\ \citenamefont
  {Sagawa}}]{PhysRevC.94.045803}%
  \BibitemOpen
  \bibfield  {author} {\bibinfo {author} {\bibfnamefont {Z.~Y.}\ \bibnamefont
  {Zhu}}, \bibinfo {author} {\bibfnamefont {A.}~\bibnamefont {Li}}, \bibinfo
  {author} {\bibfnamefont {J.~N.}\ \bibnamefont {Hu}}, \ and\ \bibinfo {author}
  {\bibfnamefont {H.}~\bibnamefont {Sagawa}},\ }\href {\doibase
  10.1103/PhysRevC.94.045803} {\bibfield  {journal} {\bibinfo  {journal} {Phys.
  Rev. C}\ }\textbf {\bibinfo {volume} {94}},\ p.\ \bibinfo {pages} {045803}
  (\bibinfo {year} {2016})}\BibitemShut {NoStop}%
\bibitem [{\citenamefont {Li}, \citenamefont {Long},\ and\ \citenamefont
  {Sedrakian}(2018)}]{Li2018}%
  \BibitemOpen
  \bibfield  {author} {\bibinfo {author} {\bibfnamefont {J.~J.}\ \bibnamefont
  {Li}}, \bibinfo {author} {\bibfnamefont {W.~H.}\ \bibnamefont {Long}}, \ and\
  \bibinfo {author} {\bibfnamefont {A.}~\bibnamefont {Sedrakian}},\ }\href
  {\doibase 10.1140/epja/i2018-12566-6} {\bibfield  {journal} {\bibinfo
  {journal} {Eur. Phys. J. A}\ }\textbf {\bibinfo {volume} {54}},\ p.\ \bibinfo
  {pages} {133} (\bibinfo {year} {2018})}\BibitemShut {NoStop}%
\bibitem [{\citenamefont {Long}\ \emph {et~al.}(2004)\citenamefont {Long},
  \citenamefont {Meng}, \citenamefont {Van~Giai},\ and\ \citenamefont
  {Zhou}}]{Long2004PRC69.034319}%
  \BibitemOpen
  \bibfield  {author} {\bibinfo {author} {\bibfnamefont {W.~H.}\ \bibnamefont
  {Long}}, \bibinfo {author} {\bibfnamefont {J.}~\bibnamefont {Meng}}, \bibinfo
  {author} {\bibfnamefont {N.}~\bibnamefont {Van~Giai}}, \ and\ \bibinfo
  {author} {\bibfnamefont {S.~G.}\ \bibnamefont {Zhou}},\ }\href@noop {}
  {\bibfield  {journal} {\bibinfo  {journal} {Phys. Rev. C}\ }\textbf {\bibinfo
  {volume} {69}},\ p.\ \bibinfo {pages} {034319} (\bibinfo {year}
  {2004})}\BibitemShut {NoStop}%
\bibitem [{\citenamefont {Typel}\ and\ \citenamefont
  {Wolter}(1999{\natexlab{b}})}]{Typel1999NPA656.331}%
  \BibitemOpen
  \bibfield  {author} {\bibinfo {author} {\bibfnamefont {S.}~\bibnamefont
  {Typel}}\ and\ \bibinfo {author} {\bibfnamefont {H.~H.}\ \bibnamefont
  {Wolter}},\ }\href@noop {} {\bibfield  {journal} {\bibinfo  {journal} {Nucl.
  Phys. A}\ }\textbf {\bibinfo {volume} {656}},\ p.\ \bibinfo {pages} {331}
  (\bibinfo {year} {1999}{\natexlab{b}})}\BibitemShut {NoStop}%
\bibitem [{\citenamefont {Todd-Rutel}\ and\ \citenamefont
  {Piekarewicz}(2005)}]{Todd2005PRL95.122501}%
  \BibitemOpen
  \bibfield  {author} {\bibinfo {author} {\bibfnamefont {B.~G.}\ \bibnamefont
  {Todd-Rutel}}\ and\ \bibinfo {author} {\bibfnamefont {J.}~\bibnamefont
  {Piekarewicz}},\ }\href@noop {} {\bibfield  {journal} {\bibinfo  {journal}
  {Phys. Rev. Lett.}\ }\textbf {\bibinfo {volume} {95}},\ p.\ \bibinfo {pages}
  {122501} (\bibinfo {year} {2005})}\BibitemShut {NoStop}%
\bibitem [{\citenamefont {Cai}\ and\ \citenamefont
  {Chen}(2012)}]{Cai2012PRC85.024302}%
  \BibitemOpen
  \bibfield  {author} {\bibinfo {author} {\bibfnamefont {B.~J.}\ \bibnamefont
  {Cai}}\ and\ \bibinfo {author} {\bibfnamefont {L.~W.}\ \bibnamefont {Chen}},\
  }\href@noop {} {\bibfield  {journal} {\bibinfo  {journal} {Phys. Rev. C}\
  }\textbf {\bibinfo {volume} {85}},\ p.\ \bibinfo {pages} {024302} (\bibinfo
  {year} {2012})}\BibitemShut {NoStop}%
\bibitem [{\citenamefont {Boguta}(1981)}]{BOGUTA1981255}%
  \BibitemOpen
  \bibfield  {author} {\bibinfo {author} {\bibfnamefont {J.}~\bibnamefont
  {Boguta}},\ }\href@noop {} {\bibfield  {journal} {\bibinfo  {journal} {Phys.
  Lett. B}\ }\textbf {\bibinfo {volume} {106}},\ p.\ \bibinfo {pages} {255}
  (\bibinfo {year} {1981})}\BibitemShut {NoStop}%
\bibitem [{\citenamefont {Lattimer}\ \emph {et~al.}(1991)\citenamefont
  {Lattimer}, \citenamefont {Pethick}, \citenamefont {Prakash},\ and\
  \citenamefont {Haensel}}]{Lattimer1991PRL66.2701}%
  \BibitemOpen
  \bibfield  {author} {\bibinfo {author} {\bibfnamefont {J.~M.}\ \bibnamefont
  {Lattimer}}, \bibinfo {author} {\bibfnamefont {C.~J.}\ \bibnamefont
  {Pethick}}, \bibinfo {author} {\bibfnamefont {M.}~\bibnamefont {Prakash}}, \
  and\ \bibinfo {author} {\bibfnamefont {P.}~\bibnamefont {Haensel}},\
  }\href@noop {} {\bibfield  {journal} {\bibinfo  {journal} {Phys. Rev. Lett.}\
  }\textbf {\bibinfo {volume} {66}},\ p.\ \bibinfo {pages} {2701} (\bibinfo
  {year} {1991})}\BibitemShut {NoStop}%
\bibitem [{\citenamefont {Kl\"ahn}\ \emph {et~al.}(2006)\citenamefont
  {Kl\"ahn}, \citenamefont {Blaschke}, \citenamefont {Typel}, \citenamefont
  {van Dalen}, \citenamefont {Faessler}, \citenamefont {Fuchs}, \citenamefont
  {Gaitanos}, \citenamefont {Grigorian}, \citenamefont {Ho}, \citenamefont
  {Kolomeitsev}, \citenamefont {Miller}, \citenamefont {R\"opke}, \citenamefont
  {Tr\"umper}, \citenamefont {Voskresensky}, \citenamefont {Weber},\ and\
  \citenamefont {Wolter}}]{Klahn2006}%
  \BibitemOpen
  \bibfield  {author} {\bibinfo {author} {\bibfnamefont {T.}~\bibnamefont
  {Kl\"ahn}}, \bibinfo {author} {\bibfnamefont {D.}~\bibnamefont {Blaschke}},
  \bibinfo {author} {\bibfnamefont {S.}~\bibnamefont {Typel}}, \bibinfo
  {author} {\bibfnamefont {E.~N.~E.}\ \bibnamefont {van Dalen}}, \bibinfo
  {author} {\bibfnamefont {A.}~\bibnamefont {Faessler}}, \bibinfo {author}
  {\bibfnamefont {C.}~\bibnamefont {Fuchs}}, \bibinfo {author} {\bibfnamefont
  {T.}~\bibnamefont {Gaitanos}}, \bibinfo {author} {\bibfnamefont
  {H.}~\bibnamefont {Grigorian}}, \bibinfo {author} {\bibfnamefont
  {A.}~\bibnamefont {Ho}}, \bibinfo {author} {\bibfnamefont {E.~E.}\
  \bibnamefont {Kolomeitsev}}, \bibinfo {author} {\bibfnamefont {M.~C.}\
  \bibnamefont {Miller}}, \bibinfo {author} {\bibfnamefont {G.}~\bibnamefont
  {R\"opke}}, \bibinfo {author} {\bibfnamefont {J.}~\bibnamefont {Tr\"umper}},
  \bibinfo {author} {\bibfnamefont {D.~N.}\ \bibnamefont {Voskresensky}},
  \bibinfo {author} {\bibfnamefont {F.}~\bibnamefont {Weber}}, \ and\ \bibinfo
  {author} {\bibfnamefont {H.~H.}\ \bibnamefont {Wolter}},\ }\href {\doibase
  10.1103/PhysRevC.74.035802} {\bibfield  {journal} {\bibinfo  {journal} {Phys.
  Rev. C}\ }\textbf {\bibinfo {volume} {74}},\ p.\ \bibinfo {pages} {035802}
  (\bibinfo {year} {2006})}\BibitemShut {NoStop}%
\bibitem [{\citenamefont {Antoniadis}\ \emph {et~al.}(2013)\citenamefont
  {Antoniadis}, \citenamefont {Freire}, \citenamefont {Wex}, \citenamefont
  {Tauris}, \citenamefont {Lynch}, \citenamefont {van Kerkwijk}, \citenamefont
  {Kramer}, \citenamefont {Bassa}, \citenamefont {Dhillon}, \citenamefont
  {Driebe}, \citenamefont {Hessels}, \citenamefont {Kaspi}, \citenamefont
  {Kondratiev}, \citenamefont {Langer}, \citenamefont {Marsh}, \citenamefont
  {McLaughlin}, \citenamefont {Pennucci}, \citenamefont {Ransom}, \citenamefont
  {Stairs}, \citenamefont {van Leeuwen}, \citenamefont {Verbiest},\ and\
  \citenamefont {Whelan}}]{Antoniadis1233232}%
  \BibitemOpen
  \bibfield  {author} {\bibinfo {author} {\bibfnamefont {J.}~\bibnamefont
  {Antoniadis}}, \bibinfo {author} {\bibfnamefont {P.~C.~C.}\ \bibnamefont
  {Freire}}, \bibinfo {author} {\bibfnamefont {N.}~\bibnamefont {Wex}},
  \bibinfo {author} {\bibfnamefont {T.~M.}\ \bibnamefont {Tauris}}, \bibinfo
  {author} {\bibfnamefont {R.~S.}\ \bibnamefont {Lynch}}, \bibinfo {author}
  {\bibfnamefont {M.~H.}\ \bibnamefont {van Kerkwijk}}, \bibinfo {author}
  {\bibfnamefont {M.}~\bibnamefont {Kramer}}, \bibinfo {author} {\bibfnamefont
  {C.}~\bibnamefont {Bassa}}, \bibinfo {author} {\bibfnamefont {V.~S.}\
  \bibnamefont {Dhillon}}, \bibinfo {author} {\bibfnamefont {T.}~\bibnamefont
  {Driebe}}, \bibinfo {author} {\bibfnamefont {J.~W.~T.}\ \bibnamefont
  {Hessels}}, \bibinfo {author} {\bibfnamefont {V.~M.}\ \bibnamefont {Kaspi}},
  \bibinfo {author} {\bibfnamefont {V.~I.}\ \bibnamefont {Kondratiev}},
  \bibinfo {author} {\bibfnamefont {N.}~\bibnamefont {Langer}}, \bibinfo
  {author} {\bibfnamefont {T.~R.}\ \bibnamefont {Marsh}}, \bibinfo {author}
  {\bibfnamefont {M.~A.}\ \bibnamefont {McLaughlin}}, \bibinfo {author}
  {\bibfnamefont {T.~T.}\ \bibnamefont {Pennucci}}, \bibinfo {author}
  {\bibfnamefont {S.~M.}\ \bibnamefont {Ransom}}, \bibinfo {author}
  {\bibfnamefont {I.~H.}\ \bibnamefont {Stairs}}, \bibinfo {author}
  {\bibfnamefont {J.}~\bibnamefont {van Leeuwen}}, \bibinfo {author}
  {\bibfnamefont {J.~P.~W.}\ \bibnamefont {Verbiest}}, \ and\ \bibinfo {author}
  {\bibfnamefont {D.~G.}\ \bibnamefont {Whelan}},\ }\href@noop {} {\bibfield
  {journal} {\bibinfo  {journal} {Science}\ }\textbf {\bibinfo {volume}
  {340}},\ p.\ \bibinfo {pages} {448} (\bibinfo {year} {2013})}\BibitemShut
  {NoStop}%
\bibitem [{\citenamefont {Baym}, \citenamefont {Pethick},\ and\ \citenamefont
  {Sutherland}(1971)}]{BPS}%
  \BibitemOpen
  \bibfield  {author} {\bibinfo {author} {\bibfnamefont {G.}~\bibnamefont
  {Baym}}, \bibinfo {author} {\bibfnamefont {C.~J.}\ \bibnamefont {Pethick}}, \
  and\ \bibinfo {author} {\bibfnamefont {P.}~\bibnamefont {Sutherland}},\
  }\href@noop {} {\bibfield  {journal} {\bibinfo  {journal} {Astrophys. J.}\
  }\textbf {\bibinfo {volume} {170}},\ p.\ \bibinfo {pages} {299} (\bibinfo
  {year} {1971})}\BibitemShut {NoStop}%
\bibitem [{\citenamefont {Baym}, \citenamefont {Bethe},\ and\ \citenamefont
  {Pethick}(1971)}]{BBP}%
  \BibitemOpen
  \bibfield  {author} {\bibinfo {author} {\bibfnamefont {G.}~\bibnamefont
  {Baym}}, \bibinfo {author} {\bibfnamefont {H.~A.}\ \bibnamefont {Bethe}}, \
  and\ \bibinfo {author} {\bibfnamefont {C.~J.}\ \bibnamefont {Pethick}},\
  }\href@noop {} {\bibfield  {journal} {\bibinfo  {journal} {Nucl. Phys. A}\
  }\textbf {\bibinfo {volume} {175}},\ p.\ \bibinfo {pages} {225} (\bibinfo
  {year} {1971})}\BibitemShut {NoStop}%
\bibitem [{\citenamefont {Oppenheimer}\ and\ \citenamefont
  {Volkoff}(1939)}]{Oppenheimer1939PR55.374}%
  \BibitemOpen
  \bibfield  {author} {\bibinfo {author} {\bibfnamefont {J.~R.}\ \bibnamefont
  {Oppenheimer}}\ and\ \bibinfo {author} {\bibfnamefont {G.~M.}\ \bibnamefont
  {Volkoff}},\ }\href@noop {} {\bibfield  {journal} {\bibinfo  {journal} {Phys.
  Rev.}\ }\textbf {\bibinfo {volume} {55}},\ p.\ \bibinfo {pages} {374}
  (\bibinfo {year} {1939})}\BibitemShut {NoStop}%
\bibitem [{\citenamefont {Tolman}(1939)}]{Tolman1939PR55.364}%
  \BibitemOpen
  \bibfield  {author} {\bibinfo {author} {\bibfnamefont {R.~C.}\ \bibnamefont
  {Tolman}},\ }\href@noop {} {\bibfield  {journal} {\bibinfo  {journal} {Phys.
  Rev.}\ }\textbf {\bibinfo {volume} {55}},\ p.\ \bibinfo {pages} {364}
  (\bibinfo {year} {1939})}\BibitemShut {NoStop}%
\bibitem [{\citenamefont {Kubis}(2007)}]{Kubis2007PRC76.025801}%
  \BibitemOpen
  \bibfield  {author} {\bibinfo {author} {\bibfnamefont {S.}~\bibnamefont
  {Kubis}},\ }\href@noop {} {\bibfield  {journal} {\bibinfo  {journal} {Phys.
  Rev. C}\ }\textbf {\bibinfo {volume} {76}},\ p.\ \bibinfo {pages} {025801}
  (\bibinfo {year} {2007})}\BibitemShut {NoStop}%
\bibitem [{\citenamefont {Lattimer}\ and\ \citenamefont
  {Prakash}(2007)}]{Lattimer2007Phyreport442.109}%
  \BibitemOpen
  \bibfield  {author} {\bibinfo {author} {\bibfnamefont {J.~M.}\ \bibnamefont
  {Lattimer}}\ and\ \bibinfo {author} {\bibfnamefont {M.}~\bibnamefont
  {Prakash}},\ }\href@noop {} {\bibfield  {journal} {\bibinfo  {journal} {Phys.
  Rep.}\ }\textbf {\bibinfo {volume} {442}},\ p.\ \bibinfo {pages} {109}
  (\bibinfo {year} {2007})}\BibitemShut {NoStop}%
\bibitem [{\citenamefont {Xu}\ \emph {et~al.}(2009)\citenamefont {Xu},
  \citenamefont {Chen}, \citenamefont {Li},\ and\ \citenamefont
  {Ma}}]{Xu2009AJ697.1549}%
  \BibitemOpen
  \bibfield  {author} {\bibinfo {author} {\bibfnamefont {J.}~\bibnamefont
  {Xu}}, \bibinfo {author} {\bibfnamefont {L.~W.}\ \bibnamefont {Chen}},
  \bibinfo {author} {\bibfnamefont {B.~A.}\ \bibnamefont {Li}}, \ and\ \bibinfo
  {author} {\bibfnamefont {H.~R.}\ \bibnamefont {Ma}},\ }\href@noop {}
  {\bibfield  {journal} {\bibinfo  {journal} {Astrophys. J.}\ }\textbf
  {\bibinfo {volume} {697}},\ p.\ \bibinfo {pages} {1549} (\bibinfo {year}
  {2009})}\BibitemShut {NoStop}%
\bibitem [{\citenamefont {Fattoyev}\ and\ \citenamefont
  {Piekarewicz}(2010)}]{Fattoyev2010PRC82.025810}%
  \BibitemOpen
  \bibfield  {author} {\bibinfo {author} {\bibfnamefont {F.~J.}\ \bibnamefont
  {Fattoyev}}\ and\ \bibinfo {author} {\bibfnamefont {J.}~\bibnamefont
  {Piekarewicz}},\ }\href@noop {} {\bibfield  {journal} {\bibinfo  {journal}
  {Phys. Rev. C}\ }\textbf {\bibinfo {volume} {82}},\ p.\ \bibinfo {pages}
  {025810} (\bibinfo {year} {2010})}\BibitemShut {NoStop}%
\bibitem [{\citenamefont {Ducoin}\ \emph {et~al.}(2011)\citenamefont {Ducoin},
  \citenamefont {Margueron}, \citenamefont {Provid\^encia},\ and\ \citenamefont
  {Vida\~na}}]{Ducoin2011PRC83.045810}%
  \BibitemOpen
  \bibfield  {author} {\bibinfo {author} {\bibfnamefont {C.}~\bibnamefont
  {Ducoin}}, \bibinfo {author} {\bibfnamefont {J.}~\bibnamefont {Margueron}},
  \bibinfo {author} {\bibfnamefont {C.}~\bibnamefont {Provid\^encia}}, \ and\
  \bibinfo {author} {\bibfnamefont {I.}~\bibnamefont {Vida\~na}},\ }\href@noop
  {} {\bibfield  {journal} {\bibinfo  {journal} {Phys. Rev. C}\ }\textbf
  {\bibinfo {volume} {83}},\ p.\ \bibinfo {pages} {045810} (\bibinfo {year}
  {2011})}\BibitemShut {NoStop}%
\bibitem [{\citenamefont {Piekarewicz}, \citenamefont {Fattoyev},\ and\
  \citenamefont {Horowitz}(2014)}]{Piekarewicz2014PRC90.015803}%
  \BibitemOpen
  \bibfield  {author} {\bibinfo {author} {\bibfnamefont {J.}~\bibnamefont
  {Piekarewicz}}, \bibinfo {author} {\bibfnamefont {F.~J.}\ \bibnamefont
  {Fattoyev}}, \ and\ \bibinfo {author} {\bibfnamefont {C.~J.}\ \bibnamefont
  {Horowitz}},\ }\href {\doibase 10.1103/PhysRevC.90.015803} {\bibfield
  {journal} {\bibinfo  {journal} {Phys. Rev. C}\ }\textbf {\bibinfo {volume}
  {90}},\ p.\ \bibinfo {pages} {015803} (\bibinfo {year} {2014})}\BibitemShut
  {NoStop}%
\bibitem [{\citenamefont {Seif}\ and\ \citenamefont
  {Basu}(2014)}]{Seif2014PRC89.028801}%
  \BibitemOpen
  \bibfield  {author} {\bibinfo {author} {\bibfnamefont {W.~M.}\ \bibnamefont
  {Seif}}\ and\ \bibinfo {author} {\bibfnamefont {D.~N.}\ \bibnamefont
  {Basu}},\ }\href@noop {} {\bibfield  {journal} {\bibinfo  {journal} {Phys.
  Rev. C}\ }\textbf {\bibinfo {volume} {89}},\ p.\ \bibinfo {pages} {028801}
  (\bibinfo {year} {2014})}\BibitemShut {NoStop}%
\bibitem [{\citenamefont {Atta}\ and\ \citenamefont
  {Basu}(2014)}]{Atta2014PRC90.035802}%
  \BibitemOpen
  \bibfield  {author} {\bibinfo {author} {\bibfnamefont {D.}~\bibnamefont
  {Atta}}\ and\ \bibinfo {author} {\bibfnamefont {D.~N.}\ \bibnamefont
  {Basu}},\ }\href@noop {} {\bibfield  {journal} {\bibinfo  {journal} {Phys.
  Rev. C}\ }\textbf {\bibinfo {volume} {90}},\ p.\ \bibinfo {pages} {035802}
  (\bibinfo {year} {2014})}\BibitemShut {NoStop}%
\bibitem [{\citenamefont {Provid{\^e}ncia}\ \emph {et~al.}(2014)\citenamefont
  {Provid{\^e}ncia}, \citenamefont {Avancini}, \citenamefont {Cavagnoli},
  \citenamefont {Chiacchiera}, \citenamefont {Ducoin}, \citenamefont {Grill},
  \citenamefont {Margueron}, \citenamefont {Menezes}, \citenamefont {Rabhi},\
  and\ \citenamefont {Vida{\~{n}}a}}]{Providencia2014}%
  \BibitemOpen
  \bibfield  {author} {\bibinfo {author} {\bibfnamefont {C.}~\bibnamefont
  {Provid{\^e}ncia}}, \bibinfo {author} {\bibfnamefont {S.~S.}\ \bibnamefont
  {Avancini}}, \bibinfo {author} {\bibfnamefont {R.}~\bibnamefont {Cavagnoli}},
  \bibinfo {author} {\bibfnamefont {S.}~\bibnamefont {Chiacchiera}}, \bibinfo
  {author} {\bibfnamefont {C.}~\bibnamefont {Ducoin}}, \bibinfo {author}
  {\bibfnamefont {F.}~\bibnamefont {Grill}}, \bibinfo {author} {\bibfnamefont
  {J.}~\bibnamefont {Margueron}}, \bibinfo {author} {\bibfnamefont {D.~P.}\
  \bibnamefont {Menezes}}, \bibinfo {author} {\bibfnamefont {A.}~\bibnamefont
  {Rabhi}}, \ and\ \bibinfo {author} {\bibfnamefont {I.}~\bibnamefont
  {Vida{\~{n}}a}},\ }\href@noop {} {\bibfield  {journal} {\bibinfo  {journal}
  {Eur. Phys. J. A}\ }\textbf {\bibinfo {volume} {50}},\ p.~\bibinfo {pages}
  {44} (\bibinfo {year} {2014})}\BibitemShut {NoStop}%
\bibitem [{\citenamefont {Lyne}\ \emph {et~al.}(2004)\citenamefont {Lyne},
  \citenamefont {Burgay}, \citenamefont {Kramer}, \citenamefont {Possenti},
  \citenamefont {Manchester}, \citenamefont {Camilo}, \citenamefont
  {McLaughlin}, \citenamefont {Lorimer}, \citenamefont
  {D{\textquoteright}Amico}, \citenamefont {Joshi}, \citenamefont {Reynolds},\
  and\ \citenamefont {Freire}}]{Lyne2004SCI303.1153}%
  \BibitemOpen
  \bibfield  {author} {\bibinfo {author} {\bibfnamefont {A.~G.}\ \bibnamefont
  {Lyne}}, \bibinfo {author} {\bibfnamefont {M.}~\bibnamefont {Burgay}},
  \bibinfo {author} {\bibfnamefont {M.}~\bibnamefont {Kramer}}, \bibinfo
  {author} {\bibfnamefont {A.}~\bibnamefont {Possenti}}, \bibinfo {author}
  {\bibfnamefont {R.}~\bibnamefont {Manchester}}, \bibinfo {author}
  {\bibfnamefont {F.}~\bibnamefont {Camilo}}, \bibinfo {author} {\bibfnamefont
  {M.~A.}\ \bibnamefont {McLaughlin}}, \bibinfo {author} {\bibfnamefont
  {D.~R.}\ \bibnamefont {Lorimer}}, \bibinfo {author} {\bibfnamefont
  {N.}~\bibnamefont {D{\textquoteright}Amico}}, \bibinfo {author}
  {\bibfnamefont {B.~C.}\ \bibnamefont {Joshi}}, \bibinfo {author}
  {\bibfnamefont {J.}~\bibnamefont {Reynolds}}, \ and\ \bibinfo {author}
  {\bibfnamefont {P.~C.~C.}\ \bibnamefont {Freire}},\ }\href {\doibase
  10.1126/science.1094645} {\bibfield  {journal} {\bibinfo  {journal}
  {Science}\ }\textbf {\bibinfo {volume} {303}},\ p.\ \bibinfo {pages} {1153}
  (\bibinfo {year} {2004})}\BibitemShut {NoStop}%
\bibitem [{\citenamefont {Kramer}\ and\ \citenamefont
  {Wex}(2009)}]{Kramer2009CQG26.073001}%
  \BibitemOpen
  \bibfield  {author} {\bibinfo {author} {\bibfnamefont {M.}~\bibnamefont
  {Kramer}}\ and\ \bibinfo {author} {\bibfnamefont {N.}~\bibnamefont {Wex}},\
  }\href {http://stacks.iop.org/0264-9381/26/i=7/a=073001} {\bibfield
  {journal} {\bibinfo  {journal} {Classical Quant. Grav.}\ }\textbf {\bibinfo
  {volume} {26}},\ p.\ \bibinfo {pages} {073001} (\bibinfo {year}
  {2009})}\BibitemShut {NoStop}%
\bibitem [{\citenamefont {{Lattimer}}\ and\ \citenamefont
  {{Schutz}}(2005)}]{Lattimer2005ApJ629979}%
  \BibitemOpen
  \bibfield  {author} {\bibinfo {author} {\bibfnamefont {J.~M.}\ \bibnamefont
  {{Lattimer}}}\ and\ \bibinfo {author} {\bibfnamefont {B.~F.}\ \bibnamefont
  {{Schutz}}},\ }\href {\doibase 10.1086/431543} {\bibfield  {journal}
  {\bibinfo  {journal} {Astrophys. J.}\ }\textbf {\bibinfo {volume} {629}},\
  p.\ \bibinfo {pages} {979} (\bibinfo {year} {2005})}\BibitemShut {NoStop}%
\bibitem [{\citenamefont {Yagi}\ and\ \citenamefont
  {Yunes}(2013)}]{Yagi2013SCI341.365}%
  \BibitemOpen
  \bibfield  {author} {\bibinfo {author} {\bibfnamefont {K.}~\bibnamefont
  {Yagi}}\ and\ \bibinfo {author} {\bibfnamefont {N.}~\bibnamefont {Yunes}},\
  }\href {\doibase 10.1126/science.1236462} {\bibfield  {journal} {\bibinfo
  {journal} {Science}\ }\textbf {\bibinfo {volume} {341}},\ p.\ \bibinfo
  {pages} {365} (\bibinfo {year} {2013})}\BibitemShut {NoStop}%
\bibitem [{\citenamefont {Link}, \citenamefont {Epstein},\ and\ \citenamefont
  {Lattimer}(1999)}]{Link1999PRL83.3362}%
  \BibitemOpen
  \bibfield  {author} {\bibinfo {author} {\bibfnamefont {B.}~\bibnamefont
  {Link}}, \bibinfo {author} {\bibfnamefont {R.~I.}\ \bibnamefont {Epstein}}, \
  and\ \bibinfo {author} {\bibfnamefont {J.~M.}\ \bibnamefont {Lattimer}},\
  }\href {\doibase 10.1103/PhysRevLett.83.3362} {\bibfield  {journal} {\bibinfo
   {journal} {Phys. Rev. Lett.}\ }\textbf {\bibinfo {volume} {83}},\ p.\
  \bibinfo {pages} {3362} (\bibinfo {year} {1999})}\BibitemShut {NoStop}%
\bibitem [{\citenamefont {Andersson}\ \emph {et~al.}(2012)\citenamefont
  {Andersson}, \citenamefont {Glampedakis}, \citenamefont {Ho},\ and\
  \citenamefont {Espinoza}}]{Andersson2012PRL109.241103}%
  \BibitemOpen
  \bibfield  {author} {\bibinfo {author} {\bibfnamefont {N.}~\bibnamefont
  {Andersson}}, \bibinfo {author} {\bibfnamefont {K.}~\bibnamefont
  {Glampedakis}}, \bibinfo {author} {\bibfnamefont {W.~C.~G.}\ \bibnamefont
  {Ho}}, \ and\ \bibinfo {author} {\bibfnamefont {C.~M.}\ \bibnamefont
  {Espinoza}},\ }\href@noop {} {\bibfield  {journal} {\bibinfo  {journal}
  {Phys. Rev. Lett.}\ }\textbf {\bibinfo {volume} {109}},\ p.\ \bibinfo {pages}
  {241103} (\bibinfo {year} {2012})}\BibitemShut {NoStop}%
\bibitem [{\citenamefont {Chamel}(2013)}]{ChamelPRL2013}%
  \BibitemOpen
  \bibfield  {author} {\bibinfo {author} {\bibfnamefont {N.}~\bibnamefont
  {Chamel}},\ }\href {\doibase 10.1103/PhysRevLett.110.011101} {\bibfield
  {journal} {\bibinfo  {journal} {Phys. Rev. Lett.}\ }\textbf {\bibinfo
  {volume} {110}},\ p.\ \bibinfo {pages} {011101} (\bibinfo {year}
  {2013})}\BibitemShut {NoStop}%
\bibitem [{\citenamefont {Baym}\ \emph {et~al.}(1969)\citenamefont {Baym},
  \citenamefont {Pethick}, \citenamefont {Pines},\ and\ \citenamefont
  {Ruderman}}]{Baym1969NAT224.872}%
  \BibitemOpen
  \bibfield  {author} {\bibinfo {author} {\bibfnamefont {G.}~\bibnamefont
  {Baym}}, \bibinfo {author} {\bibfnamefont {C.}~\bibnamefont {Pethick}},
  \bibinfo {author} {\bibfnamefont {D.}~\bibnamefont {Pines}}, \ and\ \bibinfo
  {author} {\bibfnamefont {M.}~\bibnamefont {Ruderman}},\ }\href@noop {}
  {\bibfield  {journal} {\bibinfo  {journal} {Nature}\ }\textbf {\bibinfo
  {volume} {224}},\ p.\ \bibinfo {pages} {872} (\bibinfo {year}
  {1969})}\BibitemShut {NoStop}%
\bibitem [{\citenamefont {Li}\ \emph {et~al.}(2016)\citenamefont {Li},
  \citenamefont {Dong}, \citenamefont {Wang},\ and\ \citenamefont
  {Xu}}]{LiAAstrophysicalJournal2016}%
  \BibitemOpen
  \bibfield  {author} {\bibinfo {author} {\bibfnamefont {A.}~\bibnamefont
  {Li}}, \bibinfo {author} {\bibfnamefont {J.~M.}\ \bibnamefont {Dong}},
  \bibinfo {author} {\bibfnamefont {J.~B.}\ \bibnamefont {Wang}}, \ and\
  \bibinfo {author} {\bibfnamefont {R.~X.}\ \bibnamefont {Xu}},\ }\href
  {http://stacks.iop.org/0067-0049/223/i=1/a=16} {\bibfield  {journal}
  {\bibinfo  {journal} {Astrophys. J. Suppl. S.}\ }\textbf {\bibinfo {volume}
  {223}},\ p.~\bibinfo {pages} {16} (\bibinfo {year} {2016})}\BibitemShut
  {NoStop}%
\bibitem [{\citenamefont {Watanabe}\ and\ \citenamefont
  {Pethick}(2017)}]{Watanabe2017PRL119.062701}%
  \BibitemOpen
  \bibfield  {author} {\bibinfo {author} {\bibfnamefont {G.}~\bibnamefont
  {Watanabe}}\ and\ \bibinfo {author} {\bibfnamefont {C.~J.}\ \bibnamefont
  {Pethick}},\ }\href {\doibase 10.1103/PhysRevLett.119.062701} {\bibfield
  {journal} {\bibinfo  {journal} {Phys. Rev. Lett.}\ }\textbf {\bibinfo
  {volume} {119}},\ p.\ \bibinfo {pages} {062701} (\bibinfo {year}
  {2017})}\BibitemShut {NoStop}%
\bibitem [{\citenamefont {Qian}, \citenamefont {Xing},\ and\ \citenamefont
  {Sun}(2018)}]{Qian2018}%
  \BibitemOpen
  \bibfield  {author} {\bibinfo {author} {\bibfnamefont {Z.}~\bibnamefont
  {Qian}}, \bibinfo {author} {\bibfnamefont {R.~Y.}\ \bibnamefont {Xing}}, \
  and\ \bibinfo {author} {\bibfnamefont {B.~Y.}\ \bibnamefont {Sun}},\
  }\href@noop {} {\bibfield  {journal} {\bibinfo  {journal} {Sci. China Phys.
  Mech. Astron.}\ }\textbf {\bibinfo {volume} {61}},\ p.\ \bibinfo {pages}
  {082011} (\bibinfo {year} {2018})}\BibitemShut {NoStop}%
\bibitem [{\citenamefont {Hartle}(1967)}]{Hartel1967APJ150.1005}%
  \BibitemOpen
  \bibfield  {author} {\bibinfo {author} {\bibfnamefont {J.~B.}\ \bibnamefont
  {Hartle}},\ }\href@noop {} {\bibfield  {journal} {\bibinfo  {journal}
  {Astrophys. J.}\ }\textbf {\bibinfo {volume} {150}},\ p.\ \bibinfo {pages}
  {1005} (\bibinfo {year} {1967})}\BibitemShut {NoStop}%
\bibitem [{\citenamefont {Chamel}(2012)}]{Chamel2012PRC85.035801}%
  \BibitemOpen
  \bibfield  {author} {\bibinfo {author} {\bibfnamefont {N.}~\bibnamefont
  {Chamel}},\ }\href@noop {} {\bibfield  {journal} {\bibinfo  {journal} {Phys.
  Rev. C}\ }\textbf {\bibinfo {volume} {85}},\ p.\ \bibinfo {pages} {035801}
  (\bibinfo {year} {2012})}\BibitemShut {NoStop}%
\bibitem [{\citenamefont {Hinderer}(2008)}]{Hinderer2008}%
  \BibitemOpen
  \bibfield  {author} {\bibinfo {author} {\bibfnamefont {T.}~\bibnamefont
  {Hinderer}},\ }\href {\doibase 10.1086/533487} {\bibfield  {journal}
  {\bibinfo  {journal} {Astrophys. J.}\ }\textbf {\bibinfo {volume} {677}},\
  p.\ \bibinfo {pages} {1216} (\bibinfo {year} {2008})}\BibitemShut {NoStop}%
\bibitem [{\citenamefont {Damour}\ and\ \citenamefont
  {Nagar}(2009)}]{PhysRevD.80.084035}%
  \BibitemOpen
  \bibfield  {author} {\bibinfo {author} {\bibfnamefont {T.}~\bibnamefont
  {Damour}}\ and\ \bibinfo {author} {\bibfnamefont {A.}~\bibnamefont {Nagar}},\
  }\href {\doibase 10.1103/PhysRevD.80.084035} {\bibfield  {journal} {\bibinfo
  {journal} {Phys. Rev. D}\ }\textbf {\bibinfo {volume} {80}},\ p.\ \bibinfo
  {pages} {084035} (\bibinfo {year} {2009})}\BibitemShut {NoStop}%
\bibitem [{\citenamefont {Fattoyev}, \citenamefont {Piekarewicz},\ and\
  \citenamefont {Horowitz1}(2018)}]{Fattoyev2018PRL120.172702}%
  \BibitemOpen
  \bibfield  {author} {\bibinfo {author} {\bibfnamefont {F.~J.}\ \bibnamefont
  {Fattoyev}}, \bibinfo {author} {\bibfnamefont {J.}~\bibnamefont
  {Piekarewicz}}, \ and\ \bibinfo {author} {\bibfnamefont {C.~J.}\ \bibnamefont
  {Horowitz1}},\ }\href@noop {} {\bibfield  {journal} {\bibinfo  {journal}
  {Phys. Rev. Lett.}\ }\textbf {\bibinfo {volume} {120}},\ p.\ \bibinfo {pages}
  {172702} (\bibinfo {year} {2018})}\BibitemShut {NoStop}%
\bibitem [{\citenamefont {De}\ \emph {et~al.}(2018)\citenamefont {De},
  \citenamefont {Finstad}, \citenamefont {Lattimer}, \citenamefont {Brown},
  \citenamefont {Berger},\ and\ \citenamefont
  {Biwer}}]{PhysRevLett.121.091102}%
  \BibitemOpen
  \bibfield  {author} {\bibinfo {author} {\bibfnamefont {S.}~\bibnamefont
  {De}}, \bibinfo {author} {\bibfnamefont {D.}~\bibnamefont {Finstad}},
  \bibinfo {author} {\bibfnamefont {J.~M.}\ \bibnamefont {Lattimer}}, \bibinfo
  {author} {\bibfnamefont {D.~A.}\ \bibnamefont {Brown}}, \bibinfo {author}
  {\bibfnamefont {E.}~\bibnamefont {Berger}}, \ and\ \bibinfo {author}
  {\bibfnamefont {C.~M.}\ \bibnamefont {Biwer}},\ }\href {\doibase
  10.1103/PhysRevLett.121.091102} {\bibfield  {journal} {\bibinfo  {journal}
  {Phys. Rev. Lett.}\ }\textbf {\bibinfo {volume} {121}},\ p.\ \bibinfo {pages}
  {091102} (\bibinfo {year} {2018})}\BibitemShut {NoStop}%
\bibitem [{\citenamefont {Lim}\ and\ \citenamefont
  {Holt}(2018)}]{PhysRevLett.121.062701}%
  \BibitemOpen
  \bibfield  {author} {\bibinfo {author} {\bibfnamefont {Y.}~\bibnamefont
  {Lim}}\ and\ \bibinfo {author} {\bibfnamefont {J.~W.}\ \bibnamefont {Holt}},\
  }\href {\doibase 10.1103/PhysRevLett.121.062701} {\bibfield  {journal}
  {\bibinfo  {journal} {Phys. Rev. Lett.}\ }\textbf {\bibinfo {volume} {121}},\
  p.\ \bibinfo {pages} {062701} (\bibinfo {year} {2018})}\BibitemShut {NoStop}%
\end{thebibliography}%

\end{document}